\documentclass[fleqn,usenatbib]{mnras}
\usepackage{newtxtext,newtxmath}
\usepackage[T1]{fontenc}

\DeclareRobustCommand{\VAN}[3]{#2}
\let\VANthebibliography\thebibliography
\def\thebibliography{\DeclareRobustCommand{\VAN}[3]{##3}\VANthebibliography}

\newcommand{\ha}{H~$\alpha$}
\newcommand{\pb}{Pa~$\beta$}
\newcommand{\pg}{Pa~$\gamma$}
\newcommand{\bg}{Br~$\gamma$}

\usepackage{graphicx}	
\usepackage{amsmath}	
\usepackage{pdflscape}
\usepackage{subfig}
\usepackage{microtype}
\usepackage{enumitem}
\usepackage{hyperref}
\usepackage{multirow}

\title[Hydrogen Emission from T Tauri Stars]{Hydrogen Emission from Accretion and Outflow in T Tauri Stars}

\author[T. J. G. Wilson et al.]{
T. J. G. Wilson,$^{1}$\thanks{E-mail: tjgw201@exeter.ac.uk}
S. Matt,$^{1}$
T. J. Harries,$^{1}$
G. J. Herczeg$^{2}$
\\
$^{1}$University of Exeter, Physics Building, Stocker Road, Exeter EX4 4QL, UK\\
$^{2}$Kavli Institute for Astronomy and Astrophysics, Peking University, Yiheyuan Lu 5, Haidian Qu, 100871 Beijing, People's Republic of China}
\date{Accepted XXX. Received YYY; in original form ZZZ}

\pubyear{2021}

\begin{document}
\label{firstpage}
\pagerange{\pageref{firstpage}--\pageref{lastpage}}
\maketitle

\begin{abstract}
Radiative transfer modelling offers a powerful tool for understanding the enigmatic hydrogen emission lines from T Tauri stars. This work compares optical and near-IR spectroscopy of 29 T Tauri stars with our grid of synthetic line profiles. The archival spectra, obtained with VLT/X-Shooter, provide simultaneous coverage of many optical and infrared hydrogen lines. The observations exhibit similar morphologies of line profiles seen in other studies. We used the radiative transfer code \textsc{TORUS} to create synthetic \ha{}, \pb{}, \pg{}, and \bg{} emission lines for a fiducial T Tauri model that included axisymmetric magnetospheric accretion and a polar stellar wind. The distribution of Reipurth types and line widths for the synthetic \ha{} lines is similar to the observed results. 
However, the modelled infrared lines are narrower than the observations by $\approx 80{~\rm kms}^{-1}$, and our models predict a significantly higher proportion ($\approx 90$ per cent) of inverse P-Cygni profiles. Furthermore, our radiative transfer models suggest that the frequency of P-Cygni profiles depends on the ratio of the mass loss to mass accretion rates and blue-shifted sub-continuum absorption was predicted for mass loss rates as low as $10^{-12}~M_{\odot}{\rm~ yr}^{-1}$.
We explore the effect of rotation, turbulence, and the contributions from red-shifted absorption in an attempt to explain the discrepancy in widths. Our findings show that, singularly, none of these effects is sufficient to explain the observed disparity. However, a combination of rotation, turbulence, and non-axisymmetric accretion may improve the fit of the models to the observed data.
\end{abstract}



\section{Introduction}
\label{sec:introduction}

The progenitors of Sun-like stars are the T Tauri stars. They are low mass $\lesssim 2~\textrm{M}_\odot$, young pre-main sequence stars known for their variability, strong magnetic fields, accretion disc, and mass outflows \citep{JohnsKrull:2007gn,Hartmann:2016gu}. Classical T Tauri stars (CTTS) have observed excess ultraviolet and infrared continuum; their emission lines can be strong and have complex kinematic profiles \citep{1996A&AS..120..229R}. However, the origin of the radiation is not fully understood. By modelling the T Tauri emission, insight can be gained into the accretion, outflow, and the processes of angular momentum transfer present in these stars.  

\citet{1991ApJ...370L..39K} proposed that the accretion mechanism is similar to that of neutrons stars developed by~\citet{1977ApJ...217..578G}. In this formalism, known as magnetospheric accretion, the magnetic field truncates the disc at a radius $R_{\rm T}$, where the magnetic and material stresses are of the same order. The accreting material free-falls from the disc to the stellar surface along the magnetic field, creating an accretion column or funnel. The in-falling matter thermalises the kinetic energy as it impacts the stellar surface forming a shock zone or ``hot-spots''. The large energy release heats the gas to $\approx10^{6}{\rm~K}$ and could be responsible for the high blue and UV excesses observed from T Tauri stars \citep{Hartmann:2016gu}. 
The presence of an extended magnetosphere allows for angular momentum transport, which that may help explain the observed slow T Tauri rotation \citep{2014prpl.conf..433B}. Furthermore, the magnetospheric accretion can explain the red-shifted sub-continuum absorption, known as inverse P-Cygni (IPC) profiles, seen \citep[e.g.][]{1996A&AS..120..229R,2001AA...365...90F,1994AJ....108.1056E} in some T Tauri spectra.

Complex radiative transfer models were first employed to study magnetospheric accretion in T Tauri stars by \citet{Hartmann:1994tl}. They used a two-level hydrogen model to show that magnetospheric accretion could explain the strong emission lines and blue excess of several T Tauri stars. \citet{1998ApJ...492..743M} further built on the T Tauri model by using a 2D radiative transfer code that performed a multilevel statistical equilibrium calculation for eight hydrogen levels. They concluded that the \bg{} line profiles could be reproduced for embedded objects, and the Balmer line flux agreed with observations. \citet{1998ApJ...492..743M} and \citet{2001ApJ...550..944M} constrained the accretion column temperature to be between $6000$--$10000{\rm~K}$ and determined that Stark broadening may contribute to the observed broad \ha{} wings ($\pm500{\rm~kms}^{-1}$). \citet{2007prpl.conf..479B} firmly established funnel-flow magnetospheric accretion paradigm, and it is now the generally accepted model for accretion in Classical T Tauri stars.

Blue-shifted sub-continuum absorption (the classical P-Cygni profile) is also observed and interpreted as evidence of the presence of mass outflow \citep[e.g.][]{2006ApJ...646..319E,Kwan:2006dy}. \citet{Kurosawa:2006gd} introduced a disc wind and a polar stellar winds into their T Tauri radiative transfer simulations and were able to produce all seven classical line profile classes defined by \citet{1996A&AS..120..229R} for \ha{}. \citet{Kurosawa:2011fh} showed that a bipolar stellar wind could produce a broad P-Cygni absorption, whereas a narrow P-Cygni absorption was the product of a slow disc wind. Although, it is not clear in which physical paradigms these outflow mechanisms are dominant or whether they coexist.

Another consideration for magnetospheric accretion is that the magnetic fields of T Tauri stars are rarely aligned with their poles \citep[e.g.][]{2020MNRAS.497.2142M}. A magnetic obliquity will cause the accretion funnel to favour specific longitudes of the ascending nodes preferentially \citep{2021Natur.597...41E}. A non-axisymmetric accretion funnel is likely to reduce the frequency of the observed IPC profiles, an effect that has been explored in several studies. For instance, \citet{2012A&A...541A.116A} successfully used a non-axisymmetric accretion model, based on a magnetohydrodynamic simulation, to perform a radiative transfer simulation of V2129 Oph* in \ha{} and H$\beta$. However, they were unable to reproduce the red-shifted absorption accurately.

Despite the relative success of radiative transfer studies, the T Tauri models still have multiple free parameters and chronic degeneracy. It is unclear how to disentangle the effects of geometry, accretion, and mass loss on the observed spectra. Consequently, although evidence supports the magnetospheric paradigm, the radiative transfer models need development before they can be comprehensively used as a diagnostic tool for T Tauri spectra. For example, \citet{2001AA...365...90F} noted that the current radiative transfer magnetospheric accretion models produced infrared lines (Paschen and Brackett) that were narrower than their observations by $\approx100{\rm~kms}^{-1}$.

The focus of this work is to present an initial comparison of synthetic and observed hydrogen line profiles, using the radiative transfer code \textsc{TORUS} \citep{2019A&C....27...63H}. Our primary focus is on the model development and morphology of the synthetic spectra. This paper is structured as follows: initially, we introduce the T Tauri observations and data reduction in \S~\ref{sec:observations}. We then describe in \S~\ref{sec:lineProfiles} the T Tauri model and the radiative transfer method. This section also outlines the parameters for our grid of synthetic hydrogen line profiles. The results of our parameter study, the comparison of the observations with the ensemble of synthetic profiles, and possible model modifications are presented in \S~\ref{sec:results}. Finally, we discuss the results and their implications in \S~\ref{sec:discusion} and conclude in \S~\ref{sec:conclusions}.


\section{Observations and Data Reduction}
\label{sec:observations}

\begin{table*}[htb]
 \caption{The 29 T Tauri stars used in this work. Here the notes ``B'' or ``T'' denote that the star is listed in the literature as a binary or triple system respectively, a value of the angular separation is also given where available. The complete SIMBAD name for Stars named T-\# are Ass-Cha-T-2-\#. The masses, accretion rates, and notes are adopted from: 1 \citet{2016A&A...585A.136M}, 2 \citet{2018A&A...614A.108S}, 3 \citet{2014A&A...568A..18M}, 4 \citet{2015A&A...577A..11M}, and 5 \citet{2018A&A...609A..70R}.}
 \label{tab:targetstars}
\begin{tabular}{llcccccc}
\hline
\hline
\multirow{2}{*}{Name} & \multirow{2}{*}{Type} & $T_{\rm eff}$& $A_{V}$ & Mass & $\log\dot{M}$ &\multirow{2}{*}{Notes} & \multirow{2}{*}{Ref.} \\
 & & [K] & [mag] & [${\rm~ M_{\odot}}$] & [${\rm~ M_{\odot}yr^{-1}}$] & & \\
\hline
ESO-Ha-562      & M0.5+M0.5     & 3705 & 3.4  & 0.66 & -9.3     & B 0.28"          & 1   \\
IQ-Tau          & M0.5          & 3850 & 1.7  & 0.99 & -8.2     &                  & 1   \\
V354-Mon        & K4            & 4900 &$\sim$& 1.4  & -8       &                  & 2   \\
T-33            & K0            & 5110 & 2.7  & 0.95 & -8.7     & T 2.4"           & 1   \\
T-11            & K2            & 4900 & 0.8  & 1.32 & 8.3      & B                & 3   \\
VW-Cha          & K7+M0         & 4060 & 1.9  & 1.24 & -7.9     & B 0.66"          & 1   \\
T-6             & K5            & 4350 & 0.51 & 1.22 & -7.8     &                  & 3   \\
CR-Cha          & K0            & 5110 & 1.3  & 1.78 & -8.7     &                  & 1   \\
T-52            & K0            & 5110 & 1.0  & 1.4  & -7.4     & B 11.6"          & 1   \\
T-38            & M0.5          & 3780 & 1.9  & 0.71 & -9.4     &                  & 1   \\
KV-Mon          & K4            &$\sim$&$\sim$& 1.31 & $\sim$   &                  & 4   \\
Sz-22           & K5            & 4350 & 3.2  & 1.09 & -8.4     & T 17.6"          & 1   \\
T-4             & K7            & 4060 & 0.5  & 1.03 & -9.5     &                  & 1   \\
DG-Tau          & K5            & 4350 & 2.2  & 1.4  & -7.7     &                  & 1   \\
T-23            & M4.5          & 3200 & 1.7  & 0.33 & -8.3     &                  & 1   \\
RECX-12         & M3            & 3410 &$\sim$& 0.29 & -9.8     &                  & 5   \\
Cha-Ha-6        & M6.5          & 2935 & 0.1  & 0.1  & -10.3    &                  & 1   \\
T-12            & M4.5          & 3200 & 0.8  & 0.23 & -8.8     &                  & 1   \\
CT-Cha-A        & K5.           & 4350 & 2.4  & 1.4  & -6.9     &                  & 1   \\
ESO-HA 442      & M2            &$\sim$&$\sim$& 0.36 & $\sim$   &                  & 4   \\
CHX18N          & K2            & 4900 & 0.8  & 1.17 & -8.1     &                  & 1   \\
TW-Cha          & K7.           & 4060 & 0.8  & 1    & -9       & B                & 1   \\
RECX-6          & M3            & 3415 &$\sim$& 0.15 & -11      &                  & 5   \\
T-49            & M3.5          & 3340 & 1.0  & 0.36 & -7.6     & B 24.4"          & 1   \\
T-35            & K7            & 4060 & 2.9  & 0.96 & -8.9     &                  & 3   \\
T-45a           & K7            & 4060 & 1.1  & 0.97 & -9.9     & B 28.3"          & 1   \\
RECX-9          & M4.5          & 3085 &$\sim$& 0.15 & -10.4    &                  & 5   \\
T-24            & M0            & 3850 & 1.5  & 0.91 & -8.7     &                  & 1   \\
Hn-5            & M5           & 3125 & 0.0  & 0.16 & -9.3     &                  & 1   \\
\hline
\multicolumn{6}{l}{\textbf{Note:} ESO-HA 442 is also known as CID-0223980048.}
\end{tabular}
\end{table*}

We selected 29 T Tauri stars from the ESO programme 084.C-1095(A) observed on 2010 January 18, 19, and 20 (PI G.~Herczeg, see \citealt{2016A&A...585A.136M}, \citealt{2018A&A...609A..70R}, \citealt{2018A&A...614A.108S}, and \citealt{2016ApJ...832..153L}).  The observed targets were selected to have a broad range of accretion rates of four orders of magnitude, the full list can be seen in Table~\ref{tab:targetstars}. The observations were made using VLT's X-Shooter instrument \citep{2011A&A...536A.105V} providing medium-resolution ($R~11600-18400$) spectra. The spectral range is $300-2480 {\rm~nm}$ divided into three parts, \textit{UVB}, \textit{VIS}, and \textit{NIR} corresponding to the three independent cross-dispersed echelle spectrographs on X-Shooter, which are observed simultaneously. The spectra were obtained by nodding the telescope parallel to the slit. A slit width of 0.4'' was used for all stars apart from DG Tau, which used a width of 1.2''. The spectra were flux calibrated during their initial reduction using the spectrophotometric standard star GD-71 \citet{2016A&A...585A.136M}.

Telluric corrections were applied using ESO's Molecfit pipeline \citep{2015A&A...576A..77S,2015A&A...576A..78K}. The hydrogen lines \ha{} (2-3, $6562$~\AA), \pb{} (3-5, $12818$~\AA), \pg{} (3-6, $10938$~\AA) and \bg{} (4-7, $21655$~\AA) were extracted from the full X-Shooter spectra. The profiles were normalised by fitting a third order polynomial to the nearby continuum. Veiling is expected to influence the equivalent widths of the optical and infrared lines due to excess emission from the accretion and warm dust in the disc. Although the profiles are presented in continuum normalized space, the ISM extinction will influence the unnormalised fluxes. The star's \citet{1989ApJ...345..245C} extinction law reddening corrections are given in Table~\ref{tab:targetstars}. However, the effects of veiling and extinction are not addressed in this work and are left to a future paper. Instead, the focus is on the line morphology and width, characteristics independent of the veiling and extinction.

The hydrogen line profiles from X-Shooter can be seen in Fig.~\ref{fig:herczegstars}. The figure shows the continuum normalised line profiles for the target stars arranged in order of \ha{} intensity. 
Each column displays the spectra for the individual objects listed in Table~\ref{tab:targetstars} at four different wavelengths, centred on \ha{}, \pg{}, \pb{}, and \bg{}. Some stars with strong \ha{} emission, e.g. Cha Ha 6, CID 0223980048 (ESO-HA 442), and Ass Cha T 2-33 have undetected emission in the Paschen and Brackett lines.

Following the scheme set out by \citet{1996A&AS..120..229R}, the emission-line profiles were classified as follows.
\begin{itemize}[labelindent=2pt,itemindent=0pt,leftmargin=*]
    \item \textit{Type I} symmetric profiles with little or no features.
    \item \textit{Type II} two peaks, the second being greater than half the strength of the primary.
    \item \textit{Type III} two peaks, the second being less than half the strength of the primary.
    \item \textit{Type IV} sub-continuum absorption with no further significant emission beyond the absorption. 
\end{itemize}
The types are subdivided into \textit{B} or \textit{R}, depending on if the absorption feature is blue-shifted or red-shifted relative to the primary peak. Types \textit{IVB} and \textit{IVR} are analogous to P-Cygni and IPC profiles, respectively. 

Fig.~\ref{fig:reipurth} shows the Reipurth classification for the observations and the synthetic profiles. The latter are discussed in \S~\ref{sec:results}. The line profiles were classified by eye. The seven types of profile morphology defined by \citet{1996A&AS..120..229R} are seen in the data. For spectra where the Reipurth type was not easily determined due to the noise levels, the profiles were not classified and excluded from the results of this paper. Therefore, these unclassified spectra introduce an uncertainty into the Reipurth distribution. The number of unclassified spectra are: 0 \ha{}, 10 \pb{}, 14 \pg{}, and 18 \bg{}. The sample of T Tauri stars is comparatively small and a significant proportion of the infrared lines could not be classified. Consequently, the statistical significance of the distribution is ambiguous; however, it exhibits a similar distribution of morphologies as those seen in previous T Tauri studies \citep[e.g.][]{2001AA...365...90F,1996A&AS..120..229R}. The distribution concurs with the previous results in showing that the majority ($\approx 43$ per cent) of profiles are type \textit{I} and the second highest population of Reipurth types for the infrared lines are type \textit{IVR}.

There are several cases of IPC profiles in our data, the majority of which are seen for \pb{}. There are two notable stars that contain prominent IPC signatures across all the infrared lines. Ass Cha T 2-35 has strong IPC absorption in all the infrared lines. Whereas, the \ha{} line only displays a red-shifted secondary peak. Ass Cha T 2-49 has a narrow IPC absorption in the infrared lines and no significant \ha{} red-shifted absorption. Blue-shifted absorption is even more rare in our data. For instance, Ass Cha T 2-52 has a narrow \ha{} P-Cygni absorption profile consistent with a disc wind \citep{Kurosawa:2011fh}. None of the infrared lines exhibit discernible blue-shifted absorption.

Four of the observed stars are classified as weak line T Tauri stars, according to the classification scheme of \citet{White:2003gy}, because their half widths at 10 percent maximum (HW10\%) are less than $135{\rm~kms}^{-1}$. The stars are RECX-6, Ass Cha T 2-45a, Ass Cha T 2-4, and  RECX-12. In this work, we adopt the HW10\% criterion because \citet{White:2003gy} showed it to be a reliable method of distinguishing between classical and weak line T Tauri stars and the HW10\% criterion is unaffected by veiling.

\begin{landscape}
\begin{figure}
    \centering
    \includegraphics[width=0.9\linewidth]{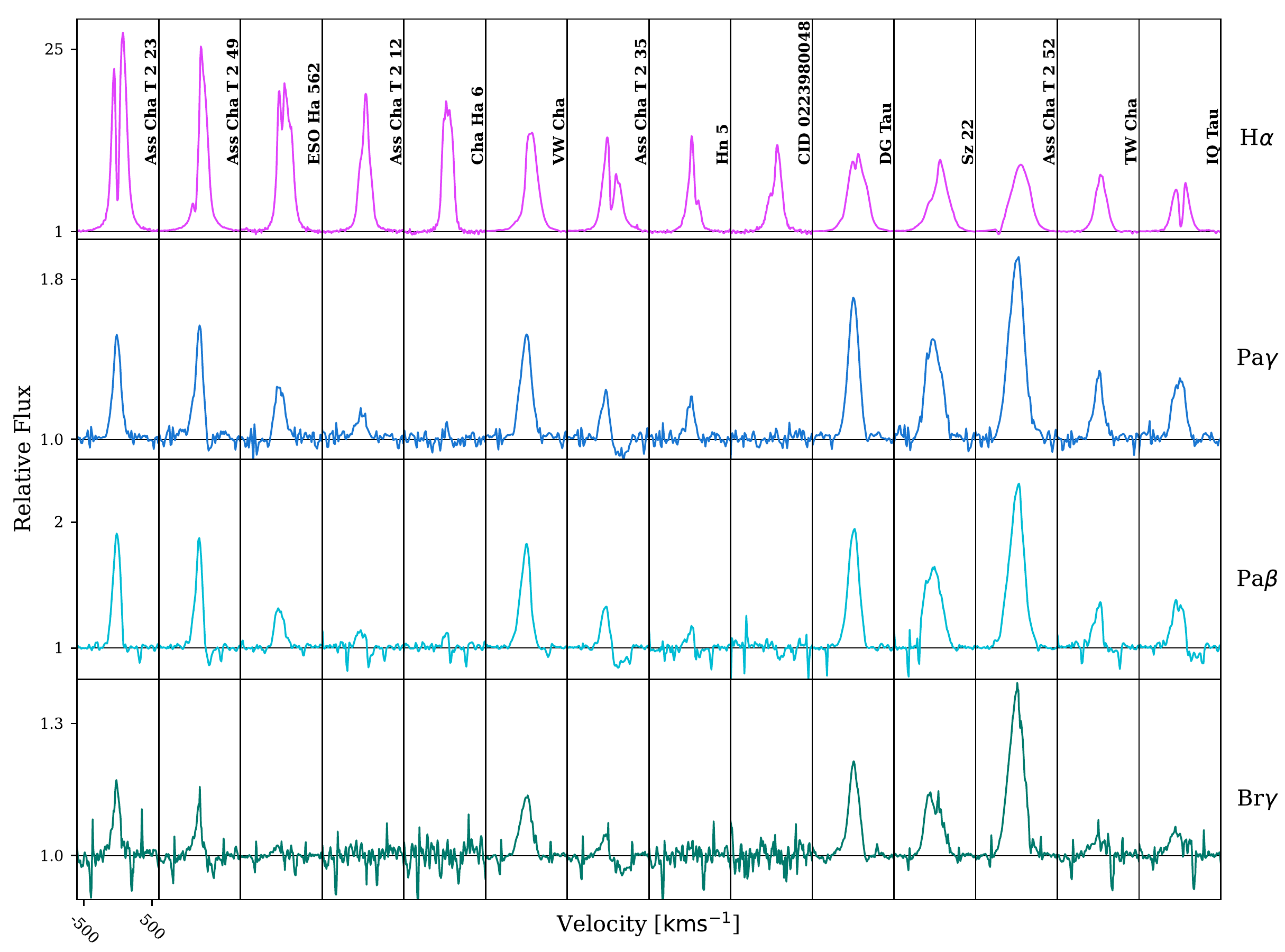}
    \caption{The normalised X-Shooter spectra of the objects (columns) listed in Table.\ref{tab:targetstars}, at four different wavelengths: \ha{}; 6562~\AA, \pg{}; 10938~\AA, \pb{}; 12818~\AA, and \bg{}; 21655~\AA. The relative flux is plotted as a function of velocity, and each column has a width of $\pm 500{\rm~ kms}^{-1}$. The stars are arranged left to right by \ha{} peak intensity.}
    \label{fig:herczegstars}
\end{figure}
\end{landscape}

\begin{landscape}
\begin{figure}
    \ContinuedFloat
    \centering
    \includegraphics[width=0.9\linewidth]{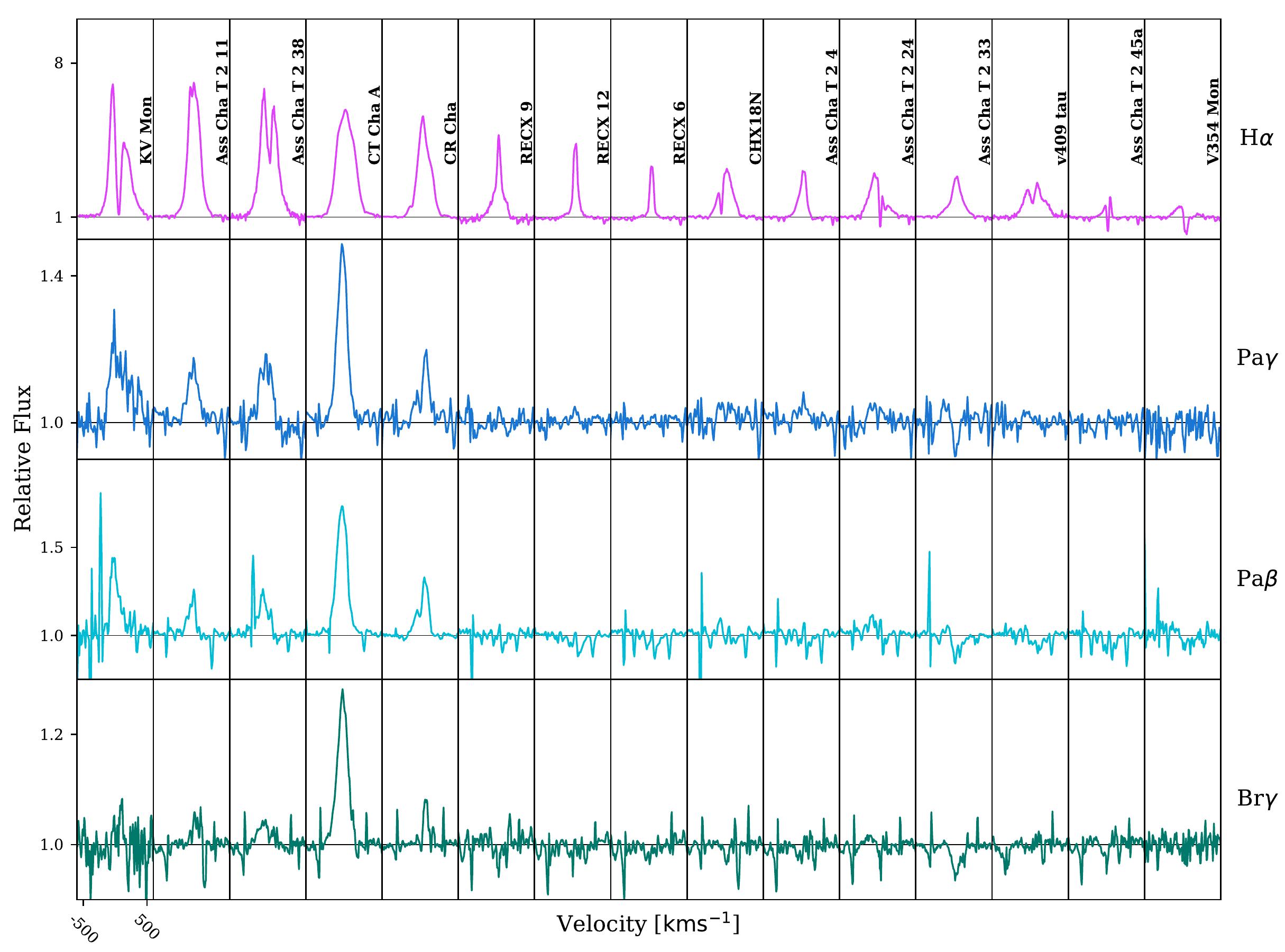}
    \caption{Continued. The stars are arranged left to right by \ha{} peak intensity. Note the different y-axis range from the previous figure.}
    \label{fig:herczegstars1}
 \end{figure}
\end{landscape}

\begin{figure}
    \centering
    \includegraphics[width=\linewidth]{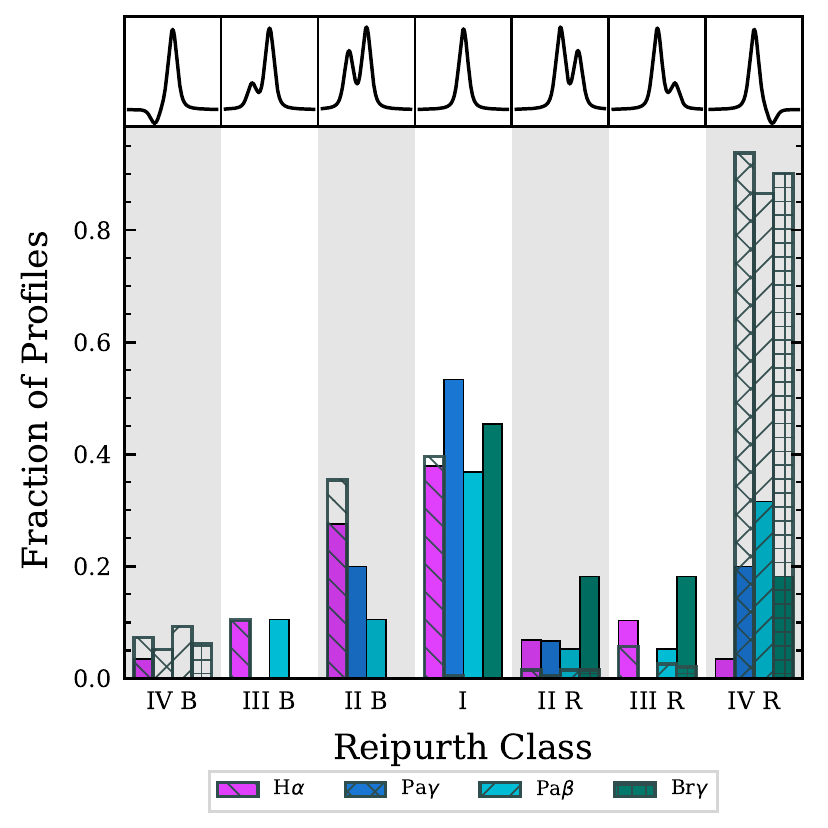} 
    \caption{Relative percentage of Reipurth classification. The coloured bars indicate the results for the X-Shooter observations. The results are subdivided by hydrogen emission line and an example profile of each type is shown at the top. The hashed bars over plotted are the classification of the corresponding synthetic emission lines.}
    \label{fig:reipurth}
\end{figure}

\section{Radiative Transfer Modelling}
\label{sec:lineProfiles}

To create synthetic T Tauri hydrogen emission lines, we used the radiative transfer code \textsc{TORUS} \citep{2019A&C....27...63H}. The code uses an analytic model as an input that approximates the expected geometry, density, temperature, velocity, emission and absorption sources in a 3D space. These include the disc, star, magnetospheric accretion, and stellar wind. \textsc{TORUS} then uses this model to calculate the expected population levels and create synthetic spectra.

\subsection{T Tauri accretion and outflow model}
\label{sec:model}

\begin{figure}
    \centering
    \includegraphics[width=\linewidth]{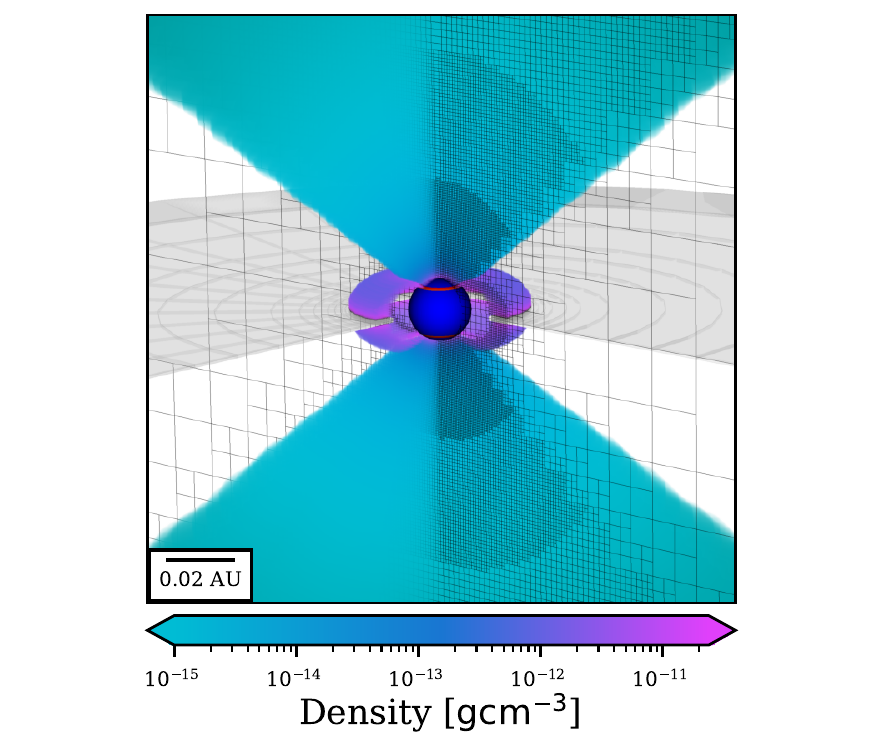}
    \caption{A 3D render of the T Tauri model from \textsc{TORUS} with a section of the stellar wind and accretion funnel cut away. The colour scale of the accretion funnel and stellar wind denote density. The hot-spot is shown on the stellar surface as a red band. The adaptive mesh (grey lines) uses a higher frequency of cells for denser regions. The disc shown is a schematic to portray the position of the geometrically thin, optically thick disc.}
    \label{fig:visitWind}
\end{figure}

\begin{figure}
    \centering
    \includegraphics[width=\linewidth]{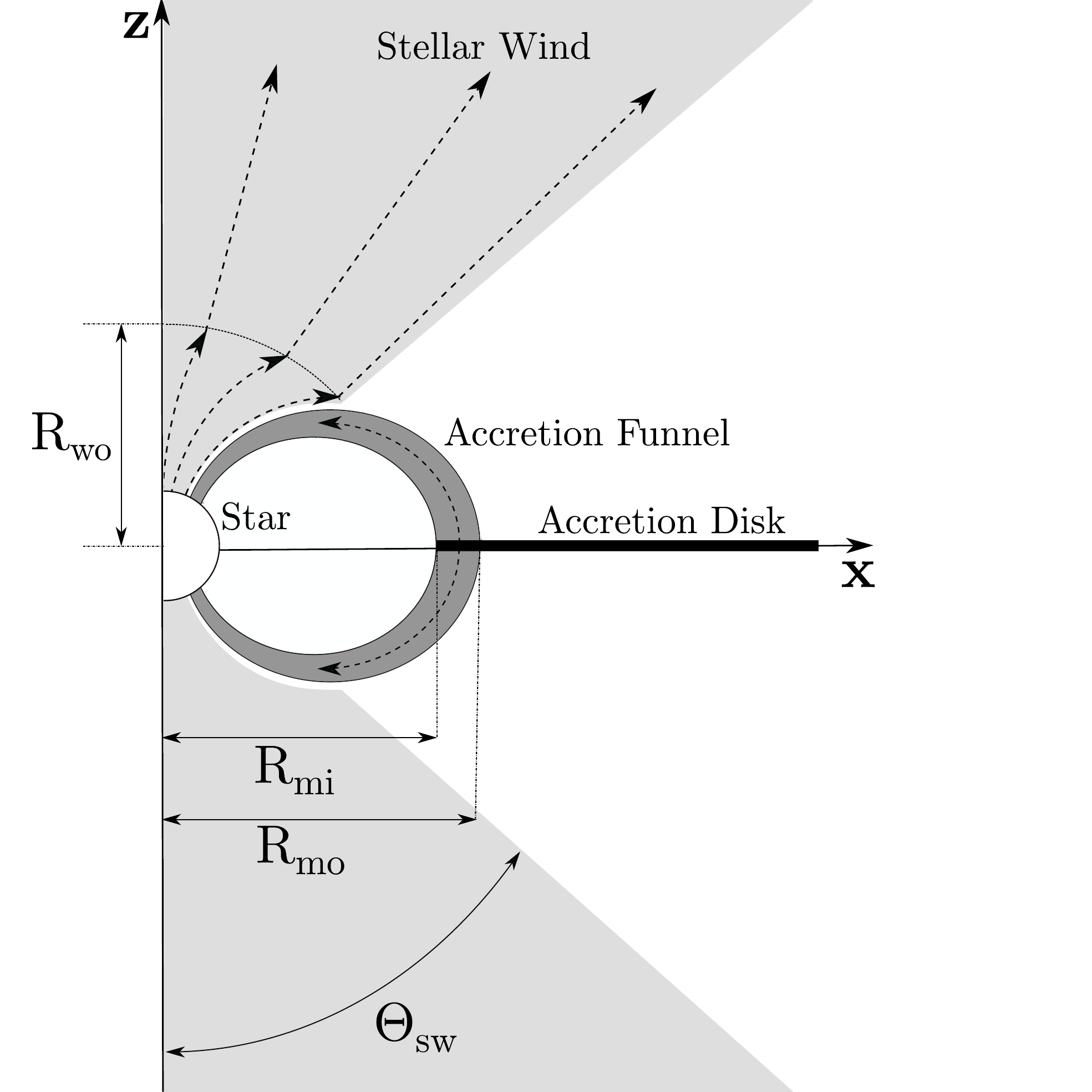}
    \caption{A cartoon depicting the T Tauri model geometry used in this paper. The models are axisymmetric around the Z-axis. The stellar wind becomes radial beyond $R_{\rm wo}$, defined as the radius at which the magnetically connected wind expands to the prescribed opening angle $\Theta_{\rm sw}$. Not to scale.}
    \label{fig:diagram}
\end{figure}

\textsc{TORUS} uses an adaptive mesh refinement (AMR) system to create a numerical grid representing the physical space of the simulation (see Fig.~\ref{fig:visitWind}). The simulations presented in this work use a 2.5D model, and they are axisymmetric around the pole. The 2.5D geometry is defined using a 2D density field and a 3D velocity field. The third component is calculated symmetrically for a given azimuthal angle.

The mesh is populated with volumes of space that are filled with a star, an accretion flow, and a stellar wind (see Fig.~\ref{fig:diagram}). An accretion disc is represented by an opaque, infinitely thin plane in the equatorial region of the grid, with a circular gap in the centre, containing the accretion flow and star. The density, temperature, and velocity fields are defined as outlined in \S\ref{sec:accretion} and \S\ref{sec:wind}. The star, placed at the origin, is defined by its radius $R_*$, mass $M_*$, and effective temperature $T_{\rm eff}$. The stellar continuum is interpolated from tabulated Kurucz model atmospheres \citep{1979ApJS...40....1K}.

\begin{table*}
 \caption{Model parameters}
 \label{tab:parameters}
 \begin{tabular}{clclc}
  \hline
  \hline
     & Parameter & Value & Unit & Description\\
  \hline
    &$M_{*}$ & $0.5$ & ${\rm M_{\odot}}$ &Star mass\\
    Star&$R_{*}$ & $2.0$ & ${\rm R_{\odot}}$ &Star radius\\
    &$T_{\rm eff}$ & $4000$ & ${\rm K}$ &Effective stellar temperature\\
   \hline
    \multirow{5}{*}{\shortstack{Accretion\\Funnel}}&$R_{\rm mi}$ & $2.2$ & $R_{*}$ & Truncation radius and inner magnetosphere connection point\\
    &$R_{\rm mo}$ & $3.0$ & $R_{*}$ & Outer connection point of magnetosphere\\
    &$T_{\rm acc}$ & $6500,7500, 8500,9500$ & ${\rm K}$ & Maximum temperature of magnetosphere\\
    &$\dot{M}_{\rm acc}$ & $10^{-7},10^{-8},10^{-9}$ & ${\rm M_{\odot}yr^{-1}}$ &Magnetospheric accretion rate\\
    &$V_{\rm rot}$&$0.0$&${\rm kms^{-1}} $&Rotation velocity\\
    \hline
    &$\Theta_{\rm sw}$ & $50$ & $\deg $ & Stellar wind opening angle\\
    &$v_{\rm min}$ & $10$ & ${\rm kms^{-1}}$ & Velocity of wind at launch\\
    Stellar &$v_{\rm max}$ & $400$ & ${\rm kms^{-1}} $ & Maximum wind velocity\\
    Wind &$\beta$ & $2.89$ &  & Wind velocity scaling\\
    &$\dot{M}_{\rm sw}$ & $0.1,0.01,0.001$ & $\dot{M}_{\rm acc}$ & Mass loss rate as a fraction of accretion rate\\
    &$T_{\rm sw}$ & $6000,8000,10000$ & ${\rm K}$ & Isothermal wind temperature\\
  \hline
    Synthetic &inc & $20,60,80$ & deg & Co-latitude of synthetic profile viewing angle\\
    Profile &$\lambda$ & $6562, 10938, 12818, 21655$ & \AA & Observed wavelengths: \ha{}, \pg{}, \pb{}, and \bg{} respectively\\
  \hline
 \end{tabular}
\end{table*}
The AMR grid starts with four cells and is subsequently refined algorithmically by splitting the cells into four subcells, such that the mass of each cell is $<10^{10}{\rm~ g}$. Each cell can be subdivided a maximum of 10 times. The total size of the simulation space was set to be a 3D cylindrical mesh with a height and radius of $400\times10^{10}{\rm~cm}$. The minimum cell size therefore has a length of $0.39\times10^{10}{\rm~cm}$.
\subsubsection{Accretion funnel}
\label{sec:accretion}

The geometry of the accretion funnel follows the prescription given by \citet{Hartmann:1994tl}. Fig.~\ref{fig:diagram} shows a cartoon representation of the geometry and Table~\ref{tab:parameters} lists adopted parameter values. The inner radius of the accretion funnel is the disc truncation radius $R_{\rm mi}$. Although MHD simulations show twisting of the magnetic fields~\citep[e.g.][]{Uzdensky:2002dg} our models only include a poloidal accretion component. The magnetosphere is assumed to be rotating as a solid body with the star. The accreting matter free-falls along the magnetic field, impacting the star's surface and creating a shock region known as the hot-spot. We assume the kinetic energy is thermalized in the star's radiating layer and that it behaves as a blackbody. The hot-spot temperature is calculated from the potential energy liberated by the accreting material. The surface of the source is divided into a grid, each with its own SED, for which a \citet{1979ApJS...40....1K} model atmosphere is used. If a source grid cell coincides with the accretion funnel, the mass flux is determined from the AMR cells directly adjacent. The mass flux $\dot{M}_{\rm cell}$, accretion velocity $v$, and cell area $A_{\rm cell}$ can be used to determine the cell temperature,
\begin{equation}
    T=\left[\frac{v^{2}\dot{M}_{\rm cell}}{2\sigma A_{\rm cell}}\right]^{\frac{1}{4}}{\rm ,}
\end{equation}
\noindent
here $\sigma$ is the Stefan–Boltzmann constant. The temperature is used to determine a blackbody SED which is added to the photospheric SED.

The accretion funnel follows the magnetic dipole streamline given by
\begin{equation}
    R = R_{0}\sin^{2}(\theta_\ast)
\end{equation}
\noindent
where $R_{0}$ is the radius at the equator of the streamline that intersect the stellar surface at colatitude $\theta_\ast$. The colatitude is the complementary angle of the latitude. The poloidal magnetic field component at radius $R$ is
\begin{equation}
    B_{\rm p}(R,\theta) = \mu R^{-3}\sqrt{4-3\sin^{2}(\theta )}
\end{equation}
\noindent
where $\mu$ is the magnetic dipole moment. Finding the unit vector of $B_p$ and converting to Cartesian coordinates gives the components of the accretion flow unit vector
\begin{equation}
     \mathbf{\hat{v}_p} = \frac{1}{\sqrt{4-3\sin^2\theta}}\left( 3\sin(\theta)\sqrt{1-\sin^2\theta}\mathbf{\hat{x}}+(2-3\sin^2\theta)\mathbf{\hat{z}}\right){\rm~.}
\end{equation}
\noindent
The magnitude of the flow at $R$ is calculated from the change in potential energy,
\begin{equation}
    v_{\rm p} = \sqrt{2GM_{\ast}\left(\frac{1}{R} - \frac{1}{R_0}\right)}{\rm .}
\end{equation}
\noindent
The velocity of the flow is $\mathbf{v_{\rm p}}=v_{\rm p}\mathbf{\hat{v}_p}$. For an accretion rate $\dot{M}_{\rm acc}$ the density at radius $R$ is given to be \citep[cf.][]{Hartmann:1994tl}
\begin{equation}
    \rho(R) = \frac{\dot{M}_{\rm acc}}{4\pi(1/R_{\rm mi} - 1/R_{\rm mo})}\frac{R^{-5/2}}{\sqrt{2GM_{\ast}}}\frac{\sqrt{4-3\sin^2\theta}}{\sqrt{1-\sin^2\theta}}{\rm~.}
\end{equation}
\noindent
The temperature distribution of the accretion funnel is less well constrained. The synthetic emission lines are sensitive to temperature variations in the accretion funnel, but the exact mechanism of heating is not known. The self-consistent thermal model produced by \citet{Martin:1996ug} does not produce hydrogen emission that is consistent with the observations \citep{1998ApJ...492..743M}, hence the temperature is provided as input parameter and the code does not solve for radiative equilibrium. The temperature profile adopted for this work is based on the distribution used by \citet{Hartmann:1994tl} who created a temperature structure using a volumetric heating rate $\propto R^{-3}$. An Alfv\'en wave flux from in-homogeneous accretion could cause this heating rate, and a schematic radiative cooling rate balances the heating. The temperature structure used in our models is interpolated and scaled from the values given in fig.~6 of \citet{Hartmann:1994tl}.

\subsubsection{Stellar wind}
\label{sec:wind}

We implemented a new stellar wind in \textsc{TORUS} that builds on the work by \citet{Kurosawa:2011fh}. Their model used a radial stellar wind launched from a spherical cap above the stellar surface so that a large opening angle could be achieved without interacting with the accretion funnel. The new wind geometry launches the outflow from the stellar surface from the polar region at higher latitudes than the accretion hot-spot. The wind and accretion funnel do not intersect because the wind follows the dipole magnetic field lines until it reaches a specified opening angle $\Theta_{\rm sw}$, at which point the wind becomes radial (see Fig.\ref{fig:diagram}). The parametrization of $\Theta_{\rm sw}$ automatically determines the radius $R_{\rm wo}$ where the wind transitions from following the dipole lines to becoming radial.  This geometry approximates the wind behaviour that is observed in MHD simulations \citep[e.g.][]{2009A&A...508.1117Z,2012MNRAS.426.2901K,2021ApJ...906....4I}. The wind is assumed to be isothermal.

The wind velocity $v_{\rm sw}$ is composed of two components the poloidal velocity $v_{\rm p}$ and the toroidal velocity $v_{\rm t}$. A beta-law is used to give the magnitude of the poloidal component,
\begin{equation}
    v_{\rm p}(R) = v_{\rm min} + \left( v_{\rm max}-v_{\rm min}\right)\left(1-\frac{R_{\ast}}{R}\right)^{\beta}
    \label{eq:beta}
\end{equation}
\noindent
where $v_{\rm min}$ and $v_{\rm max}$ are the starting and end velocities of the wind. \citet{Kurosawa:2011fh} used value of $\beta=0.5$. To obtain a physical grounding for the wind velocity, we fit equation~(\ref{eq:beta}) to the velocity profile obtained in an MHD simulation of star-disc interactions from \citet{2021ApJ...906....4I}; profile obtained by private communication L G Ireland 5 March 2020). The MHD model had similar parameters as the fiducial radiative transfer model with a truncation radius $R_{\rm mi}=2.41$. A value of $\beta=2.98$ was determined using least-squares regression. The higher $\beta$ value produces a wind that accelerates slower, creating a denser wind closer to the stellar surface. This causes a marginal narrowing of low inclinations line profiles when compared to the wind model of \citet{Kurosawa:2011fh}.

The toroidal velocity component is derived from the star's rotation rate at the colatitude of wind launch, assuming that the wind and magnetic field are perfectly coupled and that the wind corotates as a rigid body with the star out to the wind corotation radius $R_{\rm wo}$. Beyond this radius, the wind conserves angular momentum, and the toroidal component of the velocity decreases proportionally to $1/R$. The radius $R_{\rm wo}$ is the point where the wind becomes disconnected from the magnetic field such that
\begin{equation}
    R_{\rm wo} = R_{\rm mo}\sin^{2}\Theta_{\rm open}{\rm .}
\end{equation}

The density of the stellar wind is determined from the wind mass loss rate $\dot{M}_{\rm sw}$, a parameter set at run time. The wind density $\rho_{\rm sw}$ is proportional to $\dot{M}_{\rm sw}/v_{\rm sw} \cdot A(R)$, where is $A(R)$ is the surface at radius $R$ through which the wind is moving. In the range $R_* < R \leq R_{\rm wo}$ the wind is assumed to be contained within a diverging flux tube. The magnetic flux of a dipole diverges at a rate of $1/R^3$ hence the surface area is $\propto R^3$. Accordingly, the density is given by
\begin{equation}
    \rho_{\rm sw} = \frac{1}{2}\frac{\dot{M}_{\rm sw}}{v_{\rm sw}(R)\cdot A_{\ast}}\left( \frac{R_{\ast}}{R}\right)^3
\end{equation}
\noindent
where $A_{\ast}$ is the surface area of the star the wind is launched from. For radii greater than $R_{\rm wo}$, the surface $A(R)$ increases as a spherical cap (radial flow) $A(R) = 2\pi R^2 (1-\cos\Theta_{\rm open})$ so the density decreases as given by
\begin{equation}
    \rho_{\rm sw} = \frac{1}{2}\frac{\dot{M}_{\rm sw}}{v_{\rm sw}(R)\cdot A_{\rm wo}}\left( \frac{R_{\rm wo}}{R}\right)^2
\end{equation}
\noindent
where $A_{\rm wo}$ is the spherical cap with radius $R_{\rm wo}$.


\subsection{Radiative transfer}
\label{sec:radiativeTransfer}

To model the radiative emission of T Tauri stars, we use the radiative transfer code \textsc{TORUS}, which has previously been used to model the emission from T Tauri stars \citep[e.g.][]{2005MNRAS.356.1489S,Kurosawa:2006gd,Kurosawa:2011fh}. However, the treatment of the radiation field has significantly diverged from prior versions. \textsc{TORUS} now uses an atomic statistical equilibrium integration method based on the accelerated Monte Carlo scheme developed by \citet{Hogerheijde:2000wb}, in which rays are propagated to random positions in a cell in random directions. The radiative transfer equation is solved along these rays to find the local continuum radiation field. This method accelerates the convergence of the simulation in areas of significant optical depth. Additionally, \textsc{TORUS} uses full-frame co-moving ray tracing to create synthetic hydrogen line profiles. \textsc{TORUS} computes the line profiles in two stages. Firstly, it calculates the level populations assuming statistical equilibrium, and secondly, \textsc{TORUS} computes the line profiles using full-frame co-moving ray tracing. This method is known as Sobolev with exact integration (SEI) \citep{2019A&C....27...63H}

\subsubsection{Statistical equilibrium}
\label{sec:statEquil}

\textsc{TORUS} calculates the level populations assuming the Sobolev approximation, based on the method of \citet{1978ApJ...220..902K}. The populations are solved for $15$ levels with 3 more held in local thermal equilibrium (LTE). The statistical equilibrium rate equation is a balance of $R_l^{\rm R}$ the net transition from the $l^{\rm th}$ level to lower levels, the transition from the $l^{\rm th}$ level to higher levels $R_l^{\rm U}$ and the recombination and the ionization rates; $R_l^{\rm R}$ and $R_l^{\rm I}$ respectively. These components are balanced for each level such that
\begin{equation}
    R_l^{\rm L} + R_l^{\rm U} + R_l^{\rm R} - R_l^{\rm I} = 0{\rm .}
\end{equation}
\noindent
Hence, the statistical equilibrium rate equation is
\begin{multline}
    \sum_{l<u}\left[ N_{l}\left(B_{lu}\mathcal{J}_{lu} + N_{e}C_{lu}\right) - N_{u}\left(A_{ul}+B_{ul}\mathcal{J}_{lu}+N_{e}C_{ul}\right)\right]\\ + \sum_{l>u}\left[N_{l}\left(A_{lu}+B_{lu}\mathcal{J}_{lu}+N_{e}C_{lu}\right)-N_{u}\left(B_{ul}\mathcal{J}_{lu}+N_{e}C_{ul}\right)\right]\\+ N_{u}^{\ast} \left[ \int_{\nu_{u}}^{\infty} \frac{4\pi}{h\nu}a_{u}(\nu)\left( \frac{2h\nu^3}{c^2}+J_{\nu}\right)\exp{\left(-\frac{h\nu}{k_{B}T_{g}}\right)}{\rm~ d}\nu + N_{e}C_{uk}\right]\\
    -N_{u}\left[4\pi\int_{\nu_u}^{\infty} \frac{a_{u}(\nu)}{h\nu}J_{\nu}{\rm~ d}\nu +N_{e}C_{uk}\right]=0
\end{multline}
\noindent
where $A_{lu}$, $B_{lu}$ are the Einstein coefficients, $C_{lu}$ is the collision rate, $N_u$ is the level population of level $u$ in statistical equilibrium, $N^*_u$ the level population given by the Saha-Boltzmann equation for an electron density $N_{e}$ and temperature $T_{g}$, $J_\nu$ is the angle average continuum mean intensity, $a_{u}(\nu)$ refers to the photoionisation cross section of level $u$ at frequency $\nu$, and $k$ refers to the continuum state. The angle-averaged profile weighted intensity of the radiation field in the line transitions between $l$ and $u$ is $\mathcal{J}_{lu}$ and is determined using the Sobolev escape probability theory.  $\mathcal{J}_{lu}$ gives the probability that a photon is absorbed in the transition of $l\rightarrow u$ \citep{Hubeny2013} and for $l<u$ it is given by \citep[cf.][]{2019A&C....27...63H,Kurosawa:2011fh},
\begin{equation}
    \mathcal{J}_{lu}= \left(1-\beta_{lu}\right)\frac{2h\nu^{3}_{lu}}{c^{2}}\left(\frac{g_u}{g_l}\frac{N_l}{N_u}-1\right)^{-1} + \beta_{c,lu}I_{c,lu}{\rm .}
\end{equation}
\noindent
Here $g$ represents the level degeneracy, $I_{c,lu}$ the line frequency continuum intensity,  and the variables $\beta_{lu}$ and $\beta_{{\rm c},lu}$ are the Sobolev escape probabilities given by
\begin{equation}
    \beta_{lu} = \frac{1}{4\pi}\oint_{4\pi}\frac{1-{\rm~ e}^{-\tau_{lu}}}{\tau_{lu}}{\rm~ d}\Omega{\rm ,}
\end{equation}
and
\begin{equation}
    \beta_{{\rm c},lu}=\int_{\Omega_{\rm disc}}\frac{1-{\rm~ e}^{-\tau_{lu}}}{\tau_{lu}}{\rm~ d}\Omega{\rm .}
\end{equation}
\noindent
The continuum escape probability $\beta_{{\rm c},lu}$ is calculated by integrating the Sobolev optical depth $\tau_{lu}$ across the solid angle subtended by the stellar photosphere $\Omega_{\rm disc}$. The Sobolev optical depth for a velocity field $\mathbf{v}$ projected along the unit vector $\mathbf{n}$ is given by
\begin{equation}
    \tau_{lu}\left(\mathbf{n}\right)=\frac{\pi e^2}{m_{\rm e}c}\left(g_{l}f_{lu}\right)\frac{1}{v}\frac{1}{\mathbf{n}\cdot\nabla\mathbf{v}}\left(\frac{N_l}{g_l}-\frac{N_u}{g_u}\right){\rm~.}
\end{equation}
\noindent
Where $m_{\rm e}$ and $e$ are the electron mass and charge respectively and $f_{lu}$ is the oscillator strength of the line transition. Only hydrogen is assumed to be present, so the electron density is $N_{\rm e}=N\left({\rm~ H}\right)^{+}$. Thus, the conservation equation is
\begin{equation}
\sum_{n=1}^{n_{\rm max}}N_{n}+\sum_{n=n_{\rm max}}^{n_{{\rm max}}+3}N_{n}^{*}+N\left({\rm~ H}\right)^{+}=\frac{\rho}{m_{\rm H}}{\rm~.}
\end{equation}
\noindent
The above equations of statistical equilibrium are solved iteratively using a Newton-Raphson scheme for each cell, assuming the starting conditions are either an LTE solution or a set of non-LTE solutions from a nearby cell. For each cell, a set of $1024$ rays are generated that have randomly assigned frequencies, origins within the cell, and directions that are biased towards the photosphere. The total continuum intensity is summed along these rays to determine the local mean intensity. This has the advantage over the original \citet{1978ApJ...220..902K} method of geometric dilution of photospheric flux in that it can account for continuum attenuation by the intervening material in the mesh. The solution is taken to be converged when the level populations of more than $95$ per cent of the grid cells have a fractional change of less than $0.01$ per cent between iterations.


\subsubsection{Synthetic line profiles}
\label{sec:profiles}

Once the level populations have been calculated, \textsc{TORUS} can compute synthetic observations to model how the system will appear to the observer. The synthetic observations created in this work by \textsc{TORUS} are position-position-velocity (PPV) datacubes containing both spectral and spacial information. The synthetic line profiles are calculated using full-frame co-moving ray tracing, which allows for pressure broadening of the lines and is better suited to dealing with regions of high optical depth when scattering is not significant. For atomic statistical equilibrium, \textsc{TORUS} does not account for scattering and only the total intensity is considered. Pressure broadening is traditionally considered important because T Tauri Balmer lines can have wings that extend up to $\pm500{\rm~kms^{-1}}$ \citep[e.g.][]{1994AJ....108.1056E}. These broad wings are hard to achieve with model geometry alone and may be the result of Stark broadening \citep{1998ApJ...492..743M,2001ApJ...550..944M}.

\textsc{TORUS} creates the PPV datacubes using co-moving full frame ray tracing and further details can be found in \citet{2019A&C....27...63H}. To do this, a grid of bins is created, analogous to the pixels of a CCD. The number of bins and the grid's position (distance, inclination, and angle around the mid-plane) are specified parameters. From each bin, a series of rays are generated that sample the AMR mesh. The total radiation flux at a frequency $\nu$ is integrated along these rays to obtain the observed flux. \textsc{TORUS} generates multiple grids at discrete frequency intervals to build up a PPV data cube. For a ray with path $t$ the total specific intensity $I_\nu$ at a frequency $\nu$ can be defined in terms of optical depth $\tau_{\nu}$ to be,
\begin{equation}
    I_{\nu}=I_{0}{\rm~ e}^{-\tau_{\infty}}+\int_{0}^{\tau_{\infty}}S_{\nu}\left(\tau_{\nu}'\right){\rm~ e}^{-\tau_{\nu}'}{\rm~ d}\tau_{\nu}'{\rm .}
\end{equation}
\noindent
Here $I_{0}$ is the boundary intensity, $\tau_{\infty}$ is the total optical depth along the ray, and $S_{\nu}$ is the source function of the medium. The value of the initial boundary intensity $I_{0}$ depends on the path's intersection point. If the ray hits the stellar photosphere, the value is computed from the \citet{1979ApJS...40....1K} model atmosphere. If the ray intersects with the accretion hotspot the intensity is a sum of the model atmosphere intensity and the intensity given by a Plank function for a hot spot temperature of $T_{\rm hs}$. Otherwise, if the ray intersects the disc or exits the simulation space, then $I_0=0$.

The ray tracing starts from the defined data cube position-position-velocity bin (the observer) and moves across the AMR mesh until it exits the other side or hits something, for example, the photosphere or the accretion disc.
As the integration progresses along the ray, the total optical depth $\tau_{\rm total}$ between the current point and the observer is calculated. For a path segment ${\rm d}s$, the optical depth along it is 
\begin{equation}
    {\rm d}\tau_{\nu}=\int \phi_{\nu}\kappa_{\nu}{\rm d}s{\rm ,}
\end{equation}
\noindent
where $\kappa_{\nu}$ is the local transition specific absorption coefficient. $\phi_\nu$ is the line profile and is discussed further below. The change in intensity across ${\rm d}s$ is then,
\begin{equation}
    I_{\nu}^{\rm new} = I_{\nu}^{\rm old} + S_{\nu}\left(1-e^{-\tau_{\nu}}\right)e^{-\tau_{\rm total}}{\rm,}
\end{equation}
\noindent
where the source function is the ratio of the local emission $j_{\nu}$ and the absorption, 
\begin{equation}
    S_{\nu} = \frac{j_{\nu}}{\kappa_{\nu}}{\rm .}
\end{equation}
\noindent
The emission and absorption coefficients are calculated from the local level populations by
\begin{equation}
    j_\nu = \frac{h}{4\pi} \cdot A \nu_{lu}N_{u}
\end{equation}
\noindent
and
\begin{equation}
  \kappa_\nu = \frac{h}{4\pi}\cdot\left(N_{l}B_{lu}-N_{u}B_{ul}\right){\rm ,}
\end{equation}
\noindent
where $A$ and $B$ are the Einstein coefficients.

Initially, each cell of the AMR mesh is divided into two ray segments, but this number is doubled if ${\rm d}\tau_\nu > 0.1$ and ${\rm d}\tau_\nu < 20$. Furthermore, if the velocity gradient across the cell would cause the line resonance to be traversed or $\nu$ is close to line resonance the number of segments is set to $20$.

The emission and absorption coefficients are affected by the local velocity, as a change in relative velocity could Doppler shift the line into resonance. This effect is contained within the line profile function $\phi_\nu$. If the intrinsic line broadening is negligible, then the profile is defined by thermal Doppler broadening and $\phi_{\nu}$ is a Gaussian profile. For optically thick lines such as \ha{}, pressure broadening becomes important and a normalised Voigt profile is adopted, which is a convolution of a Gaussian and a Lorentzian  \citep[cf.][]{1973ApJ...184..605V}. The Voigt profile is given as $\phi_{\nu}=\pi^{-1/2}H\left(a,y\right)$ where
\begin{equation}
    H\left(a,y\right) \equiv \frac{a}{\pi}\int_{-\infty}^{\infty} \frac{{\rm~ e}^{-y'^{2}}}{\left(y-y'\right)^{2}+a^{2}}{\rm~ d}y'{\rm~~.}
\end{equation}
\noindent
Here $y=\left(\nu-\nu_0\right)/{\Delta\nu_{\rm D}}$, $y'=\left(\nu'-\nu_0\right)/{\Delta\nu_{\rm D}}$, and $a=\tfrac{\Gamma}{4\pi\Delta\nu_{\rm D}}$ where $\nu_0$ is the line centre frequency. 
$\Delta\nu_{\rm D}$ is the Doppler line width of hydrogen given by
\begin{equation}
    \Delta\nu_{\rm D}=\frac{\nu_{0}}{c}\sqrt{\frac{2Tk_{\rm B}}{m_{\rm H}}+v^{2}_{\rm turb}}
    \label{eq:doppler}
\end{equation}
\noindent
where $m_{H}$ is the mass of hydrogen, $k_{\rm B}$ is the Boltzmann constant, and $v^{2}_{\rm turb}$ is the turbulent broadening factor \citep{2017SSRv..210..109D}, set to be zero for the main grid. The damping constant $\Gamma$ is as given by \citet{Luttermoser:uj},
\begin{equation}
    \Gamma = C_{\rm rad} + C_{\rm vdW}\left(\frac{N_{\rm H\,\textsc{I}}}{10^{16}}\right)\left(\frac{T}{5000}\right)^{0.3}+C_{\rm stark}\left(\frac{N_{\rm e}}{10^{12}}\right)
    \label{eq:starDamp}
\end{equation}
\noindent
where $C_{\rm rad}$, $C_{\rm vdW}$, and $C_{\rm stark}$ are the radiative, van der Waals, and Stark broadening half-widths. $N_{\rm H\,\textsc{I}}$ is the neutral hydrogen density and $T$ is the gas temperature in Kelvin. For \ha{} the following broadening parameters were adopted from \citet{Luttermoser:uj} $C_{\rm rad}=8.2\times10^{-3}$~\AA, $C_{\rm vdW}=5.5\times10^{-3}$~\AA, and $C_{\rm stark}=1.47\times10^{-2}$~\AA. Because hydrogen exhibits linear Stark broadening \citep{1964plsp.book.....G,Vernazza:1973vq}, in this work we use the method laid out by \citet{Luttermoser:uj,Kurosawa:2011fh} and \citet{1981ApJS...45..635V}.

The Stark broadening has a significant impact on the \ha{} line widths at high accretion  rates and funnel flow temperatures, see \S~\ref{sec:validation}. By using half widths determined from the extended VCS tables \citep{Lemke:1997fj}, we found that for \pg{}, \pb{} and \bg{} Stark broadening had a negligible effect.

By integrating the total surface brightness in each of the PPV datacube's frames, the synthetic line profile of the flux versus velocity are calculated. The flux is normalised relative to the first velocity bin of the datacube, which is assumed to be the continuum. This is correct as long as the full extent of the line profile is covered by the wavelength range of the datacube and the initial frame does not have any line emission. The synthetic continuum includes emission from the stellar photosphere and accretion hot-spot. The emission from dust is not included in these models. \textsc{TORUS} can run a Monte Carlo radiative transfer loop, along side the atomic statistical equilibrium calculation to determine the emission from dust; however, this is left to a future paper.
\subsection{Model validation}
\label{sec:validation}

To validate the new radiative transfer routines implemented here, \textsc{TORUS} was used to reproduce the simple \ha{} magnetospheric only model presented in fig.~3 of \citet{Kurosawa:2006gd}. The authors used a  non-linear Stark broadening coefficient defined as
\begin{equation}
    \Gamma = C_{\rm rad} + C_{\rm vdW}\left(\frac{N_{HI}}{10^{16}}\right)\left(\frac{T}{5000}\right)^{0.3}+C_{\rm stark}\left(\frac{N_e}{10^{12}}\right)^{2/3}{\rm .}
    \label{eq:starDampKuro}
\end{equation}
\noindent
The model used in fig.~3 of \citet{Kurosawa:2006gd} is identical to the setup described in \S~\ref{sec:model}, but with no stellar wind. A grid of models were run with three different accretion rates of $10^{-7},10^{-8},10^{-9}{~\rm M}_{\odot}{\rm yr}^{-1}$ and four separate accretion maximum temperatures of $6500,7500,8500,9500 {~\rm K}$. The line profiles were computed for a viewing inclination of $55^\circ$. All other parameters are identical to those presented in Table~\ref{tab:parameters}. The stark broadening parameters are the same as those presented in \S~\ref{sec:profiles}. In other words, the same fundamental physics is used by both \citet{Kurosawa:2006gd} and \textsc{TORUS}. This comparison therefore highlights the influence of the upgrades to \textsc{TORUS} (see \S~\ref{sec:radiativeTransfer}).

\begin{figure}
    \centering
    \includegraphics[width=\linewidth]{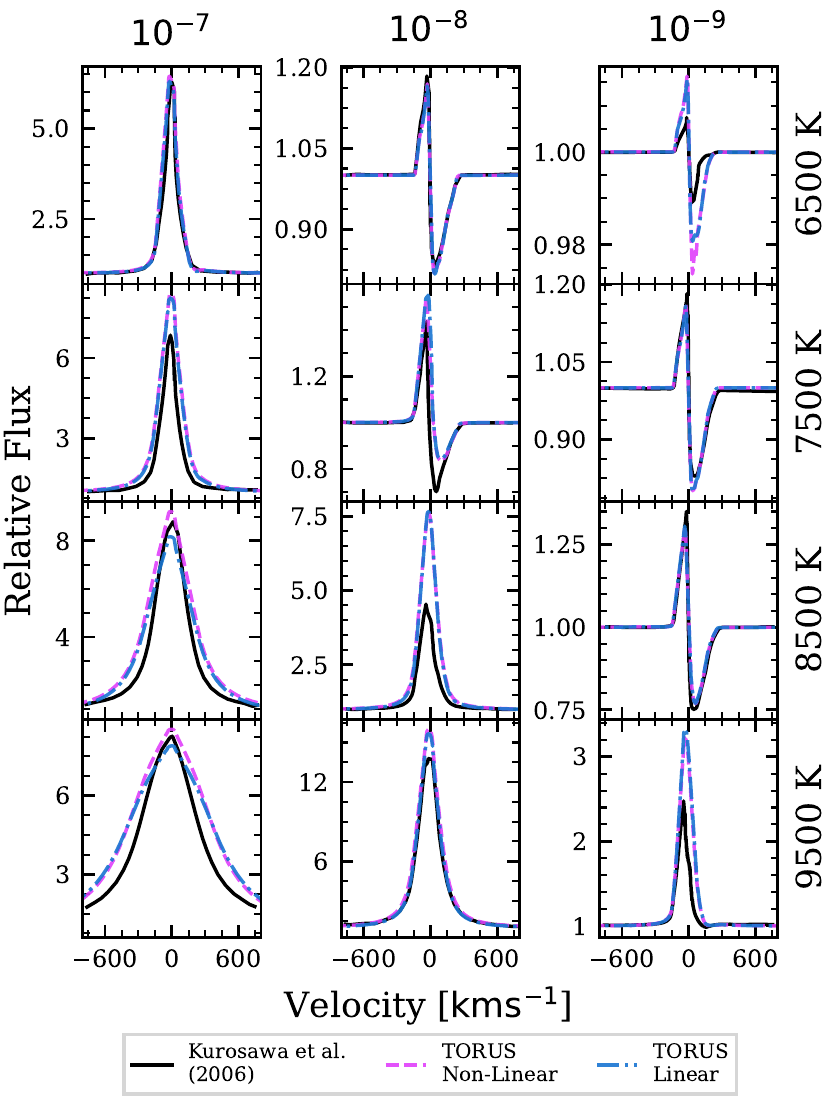}
    \caption{A comparison of synthetic \ha{} line profiles produced by \citet{Kurosawa:2006gd} (black lines) and \textsc{TORUS}. The models have magnetospheric accretion only and no form of mass outflows.  Profiles arrange by accretion rate (columns) in values of ${\rm M}_{\odot}{\rm yr}^{-1}$ and accretion maximum temperature (rows) for a viewing colatitude of $55^{\circ}$. Two different Stark broadening parametrizations are used by \textsc{TORUS}; linear (blue dot-dashed) and  non-linear (pink dashed) Stark broadening. Line strength increases with temperature and accretion rate, top-right to bottom-left.}
    \label{fig:kurosawa2006}
\end{figure}

A comparison of the results produced by \citet{Kurosawa:2006gd} and \textsc{TORUS} can be seen in Fig.~\ref{fig:kurosawa2006}. We ran the model on \textsc{TORUS} using both linear and  non-linear Stark broadening. Fig.~\ref{fig:kurosawa2006} demonstrates that the new radiative transfer scheme produces results in satisfactory agreement with the models of \citet{Kurosawa:2006gd}. The \ha{} lines produced using \textsc{TORUS} are slightly broader, with higher peak intensities than those presented by \citet{Kurosawa:2006gd}. There are only negligible differences between the lines computed with linear and  non-linear Stark broadening.

The general line profile trends shown by \citet{Kurosawa:2006gd} are reflected by our \textsc{TORUS} models. The line strength and width increase with the accretion temperature and rates. An exception to this trend is seen in Fig.~\ref{fig:kurosawa2006} for the line profile produced with an accretion rate of $\log{\dot{M}_{\rm acc}}=-8$ and $T_{\rm acc}=9500{\rm~K}$. The line profile has a greater relative peak flux than the profile produced with an accretion rate of $\log{\dot{M}_{\rm acc}}=-7$. This is because the continuum level from the hotspot is increased by the greater accretion rate, reducing the peak flux relative to the continuum. We see, in Fig.~\ref{fig:kurosawa2006}, that the width of the profiles increases with the rate and temperature of the accretion; it is unaffected by the continuum level.

The above comparisons did not include any outflows. Therefore, to test our stellar wind geometry (\S~\ref{sec:wind}) we ran an additional test to compare our model with the stellar wind geometry used by \citet{Kurosawa:2011fh}. These results are not shown but they are described below. \textsc{TORUS} was used to run two models with identical stellar and magnetospheric parameters as those defined for fig.~10 of \citet{Kurosawa:2011fh}. The first models had an identical wind as \citet{Kurosawa:2011fh}; the wind launched from $3{\rm R}_\ast$ above the star's surface. The second model used our wind geometry with the outflow being launched from the star's surface. We adopted the same outflow parameters of $T_{\rm sw}$, $\dot{M}_{\rm sw}$, $v_{\rm min}$, $v_{\rm max}$, and $\Theta_{\rm open}$. The second model also used a value of $\beta=1.6$ rather than $\beta=0.5$ as this more closely matched the wind acceleration and velocity at radii greater than $3 {\rm R}_\ast$.

The differences in the line profiles produced by the two wind models are modest, comparable in magnitude to the variations seen in Fig.~\ref{fig:kurosawa2006}. For \ha{} at low colatitude, the lines produced by our stellar wind model were broader at the peak by $\approx 20{\rm~ ~kms}^{-1}$. For high colatitudes the lines were slightly narrower and exhibited small-scale blue-shifted absorption features not seen in the \citet{Kurosawa:2011fh} wind model. For the higher lines, at inclinations of $20^{\circ}$ and $50^{\circ}$, the wind that extended to the star's surface produced lines that were narrower and had a lower peak flux. Our wind model also produced deeper red-shifted absorption than the model of \citet{Kurosawa:2011fh}. These effects, however, are likely to not be discernible in T Tauri observations. Nevertheless, the new wind geometry was used in this work because it is more physically realistic.


\subsection{Parameter study}
\label{sec:parameterStudy}

We present a grid of synthetic models covering a broad range of temperatures and accretion rates, with the parameters shown in Table~\ref{tab:parameters}. The models have a single source star with an effective temperature of $T_{\rm eff}=4000{\rm~ K}$, a mass of $M_{*}=0.5{\rm~ M}_{\odot}$, and a radius of $R_{*}=2.0{\rm~ R}_{\odot}$. Our grid covers a broad range of accretion rates of $10^{-7}$, $10^{-8}$ and $10^{-9}{\rm~ M_{\odot}yr}^{-1}$ with mass-loss rates of $0.1$, $0.01$ and $0.001$ of the accretion rate. These mass loss rates were selected to be in the required range to provide a significant `spin-down' torque on the star \citep{2005ApJ...632L.135M} and are consistent with the range of values estimated from jets \citep{2018A&A...609A..87N}. The source star parameters and the wind and magnetospheric geometries were not varied. For each model configuration, synthetic line profiles were computed for three inclinations ($20^{\circ},60^{\circ},80^{\circ}$) centred on the four hydrogen emission lines of the observational data set described in \S~\ref{sec:observations}.

\begin{landscape}
\begin{figure}
    \centering
    \includegraphics[width=0.85\linewidth]{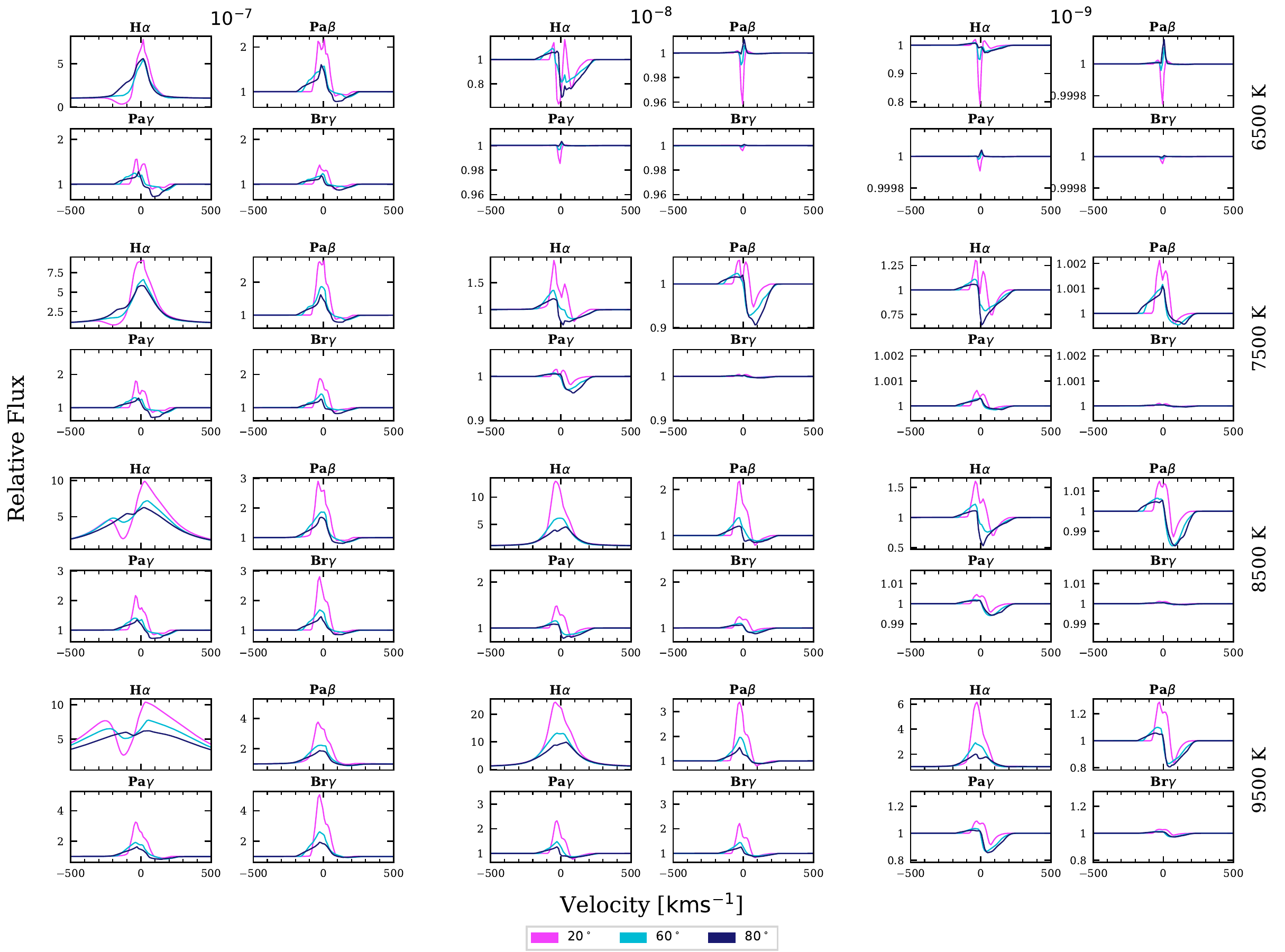}
    \caption{A sample of the line profiles produced by \textsc{TORUS}. The spectra are shown for $\dot{M}_{\rm sw}=0.01~\dot{M}_{\rm acc}$ and $T_{\rm sw}=8000{\rm~K}$. The lines are arranged by $\dot{M}_{\rm acc}$ (columns) and $T_{\rm acc}$ (rows). Each section is subdivided by hydrogen transition and shows the three different inclinations (colours). The y-axes of \pg{}, \pb{}, and \bg{} for each set, share a common range. IPC profiles are present in the majority of the infrared lines. For the hydrogen transitions at an inclination of $80^{\circ}$ (outside of the stellar wind opening angle), there is less blue-shifted absorption and no P-Cygni features. \ha{} profiles are broad with some wings exceeding $500{\rm~ kms}^{-1}$.}
    \label{fig:allModels}
\end{figure}
\end{landscape}

Fig.~\ref{fig:allModels} shows a subset of line profiles in the grid; only models with a mass loss rate of $0.01~\dot{M}_{\rm acc}$ and wind temperature of $8000{\rm~K}$ are included. The figure arranges the line profiles by accretion rate and maximum magnetospheric temperature, subdivided by the hydrogen transition. The synthetic lines exhibit a broad range of different morphologies. \ha{} has the highest peak intensity, and the subsequent strongest lines are \pb{}, \pg{}, and \bg{} respectively. However, $\approx7$ per cent of the \bg{} lines at an inclination of $20^{\circ}$ have a greater peak intensity than \pb{} and \pg{}. These lines occur for an accretion temperature of $9500{~\rm K}$ and an accretion rate of $10^{-7}{~\rm M}_{\odot}{\rm yr}^{-1}$. The \ha{} lines are in general broader than the infrared lines, with a few having broad wings $>1000{\rm~ kms}^{-1}$. The majority ($\approx83$ per cent) of the infrared lines display red-shifted absorption from the magnetosphere. Blue-shifted sub-continuum absorption is not seen at high inclination where the viewing angle is along the disc outside of the wind opening angle. 

\section{Results}
\label{sec:results}

The general trends seen in the grid of synthetic line profiles are outlined as follows. Higher accretion rates increase the line width and intensity. A high accretion rate of $\log{\dot{M}_{\rm acc}}=-6$ produced structured \ha{} absorption superposed on broad ($\pm1000{\rm~kms}^{-1}$) emission profiles. Only a limited number of models with an accretion rate of $\log{\dot{M}_{\rm acc}}=-6$ were run and the profiles are not included in this paper. The line profile intensity increases with magnetospheric temperature. The \ha{} lines are dominated by Stark broadening for accretion flow temperatures of at least $7500{\rm~K}$ at accretion rates of $\log{\dot{M}_{\rm acc}}\geq-8$. The Stark broadening significantly increases the line widths and is highly dependent on the magnetospheric temperature and density. For a temperature of $9500{\rm~K}$ and an accretion rate of $\log{\dot{M}_{\rm acc}}=-7$, the \ha{} profiles are Stark broadened to have wings of over $\pm 1000{\rm~kms}^{-1}$. Conversely, the infrared lines are almost completely unaffected by Stark broadening. The intensity and equivalent width of the infrared lines increase significantly with greater accretion rates and temperature. Whereas, the full widths at half maximum (FWHM) of the profiles are less strongly affected by the rate and temperature of the accretion flow. 

\ha{} exhibits IPC profiles for accretion rates of $\log{\dot{M_{\rm acc}}}=-8$ when $T_{\rm acc} \leq 7500{~\rm K}$ and for $\log{\dot{M}_{\rm acc}}=-9$ when $T_{\rm acc} \leq 8500{~\rm K}$. For accretion rates and temperatures above this, the Stark broadening smooths over the absorption features. For the infrared lines, the IPC profiles are exhibited irrespective of inclination, accretion rate, or temperature.

Strong P-Cygni profiles are seen when the wind outflow rate is $\geq0.01\dot{M}_{\rm acc}$ and the inclination is within the stellar wind opening angle. The continuum relative depth of the blue-shifted absorption increases with the wind temperature. The line intensity is inversely correlated with the wind temperatures. This is particularly apparent for the infrared lines, because the higher energy levels, 5, 6, and 7, become depopulated before the upper \ha{} level. However, this inverse relationship does not hold at a high wind temperature ($2\times 10^{4}{\rm~K}$), where emission from the wind dominates when the mass loss rate is $\geq0.01\dot{M}_{\rm acc}$. 

The frequency of P-Cygni profiles is correlated with the ratio $\dot{M}_{\rm sw}/\dot{M}_{\rm acc}$ and not $\dot{M}_{\rm sw}$ alone. For example, at a high accretion rate of $\log{\dot{M}_{\rm acc}}=-7$, P-Cygni profiles are only seen at mass-loss rates of $\log{\dot{M}_{\rm sw}}=-8$ and $-9$. Whereas, for an acretion rate of $\log{\dot{M}_{\rm acc}}=-8$, P-Cygni profiles are seen for mass-loss rates of $\log{\dot{M}_{\rm sw}}=-9$, $-10$ and $-11$ with decreasing frequency. Similar results are seen for an accretion rate of $\log{\dot{M}_{\rm acc}}=-9$. P-Cygni profiles were produced for mass-loss rates as low as $10^{-12}~{\rm M_{\odot}yr}^{-1}$. At higher accretion rates, the emission from the magnetosphere more commonly dominates, overwhelming the blue-shifted absorption.

The synthetic line profiles were calculated for velocity increments of $\Delta v=12.04{\rm~kms}^{-1}$, which corresponds to a spectral resolution of $R=24900$. The observations are observed at resolutions of $R=18400$ ($16{\rm~kms}^{-1}$) for \ha{} and $R=11600$ ($26{\rm~kms}^{-1}$) for the infrared lines. We explored the effect of applying a Gaussian convolution to the synthetic spectra to reduce their resolution to the resolution of the observed spectra. However, the change in line profile morphology, in particular the FWHM and HW10\%, was negligible when the synthetic spectra were convolved. Therefore, convolution was not considered a significant factor and not included in our analysis.

\begin{figure}
    \centering
    \includegraphics[width=\linewidth]{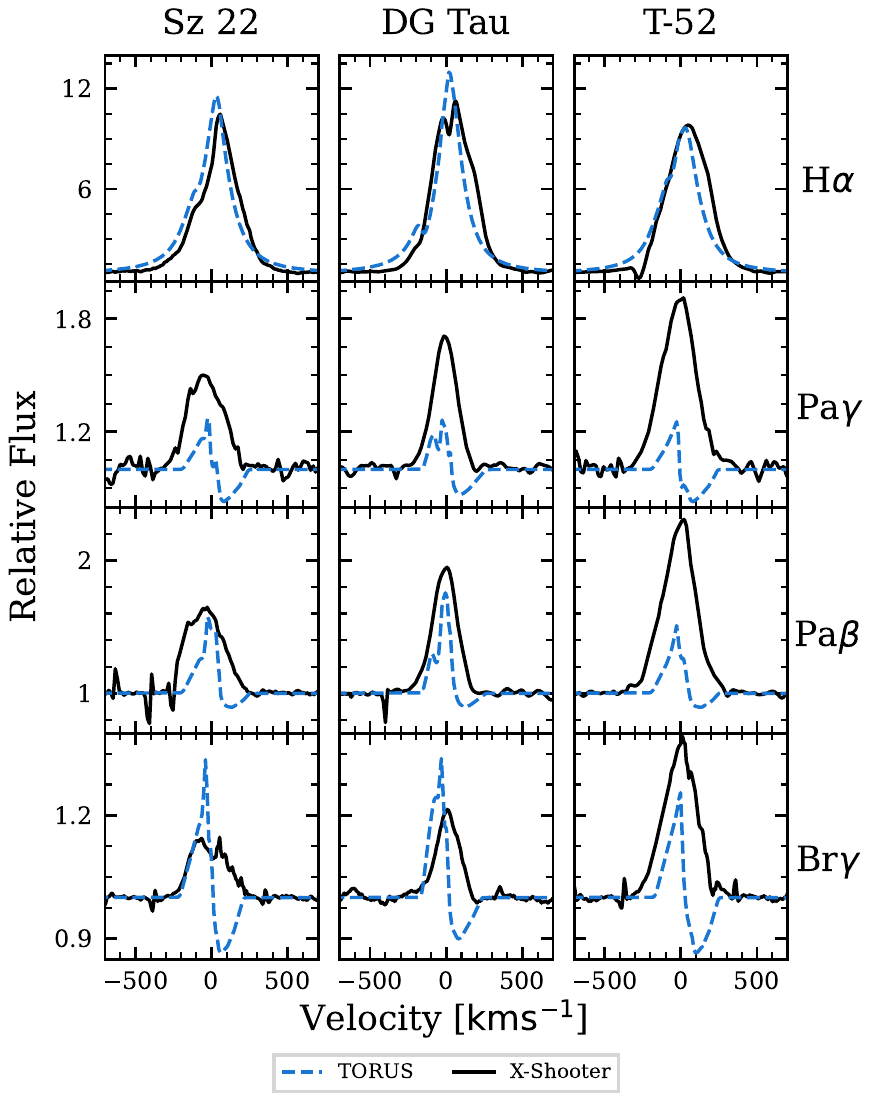}
    \caption{Three of the best cases of the grid matching the observed line profiles. The models have been chosen because they best reproduce the observed \ha{} shape and width. The observed spectra have been adjusted to compensate for each system's radial velocity.}
    \label{fig:exampleFit}
\end{figure}

Although, our model grid was not set up to fit any particular star, it is informative to directly compare the synthetic line profiles to the observed line profiles. As an example, we show the spectra of three stars (Sz 22, DG Tau, and Ass Cha T 2-52) in Fig.~\ref{fig:exampleFit} and a corresponding model selected to match the \ha{} line profiles. The models were chosen using a $\chi^{2}$ fit test to find the closest \ha{} profiles from the grid. The parameters of the chosen models are shown in Table~\ref{tab:examples}. The models use our fiducial values of $T_{\rm eff}=4000{\rm~K}$ and $M_{\ast}=0.5{\rm M}_{\odot}$, whereas the effective photospheric temperature for Sz 22 is $4350{\rm~K}$, for DG Tau it is $4350{\rm~K}$, and for Ass Cha T 2-52 it is $5110{\rm~K}$ \citep{2016A&A...585A.136M,2012ApJ...745..119N}. The stellar masses for Sz 22, DG Tau and Ass Cha T 2-52 are $1.09$, $1.4$, and $1.4{\rm~M}_{\odot}$, respectively \citep{2016A&A...585A.136M}.

A caveat to the direct comparison is that the veiling of the observed lines have not been considered. Whereas, the effect of reddening is negated by the continuum normalization, the influence of veiling is not. Nevertheless, Fig.~\ref{fig:exampleFit} highlights a trend seen in the grid of synthetic spectra. Namely, that while our method selects models with similar strengths and widths in \ha{} to that observed, the models fail to reproduce the observed infrared lines. The synthetic infrared lines are too structured and narrow. This trend is presented in more detail in \S~\ref{sec:widths} and \S~\ref{sec:linebroadening}, and discussed in \S~\ref{sec:discusion}.

In the following sections, we present an analysis of the structure (\S~\ref{sec:reipurth}) and the width (\S~\ref{sec:widths}) of the ensemble of synthetic line profiles. To best compare the synthetic and observed spectra, we use only a subset of our grid. The models are selected to correspond to the observed \ha{} range. Only models that have an \ha{} HW10\% in the range of $60-400{\rm~kms}^{-1}$ are included.

\begin{table}
\centering
\caption{Parameters of the models presented in Fig.~\ref{fig:exampleFit}.}
\begin{tabular}{llll}
\hline
\hline
Object                    & Sz 22        & DG Tau       & T-52         \\
\hline
$\log{\dot{M}_{\rm acc}}$ & $-8$         & $-8$         & $-8$         \\
$T_{\rm acc}$             & $9500$K      & $9500$K      & $9500$K      \\
$\log{\dot{M}_{\rm sw}}$      & $-9$         & $-9$         & $-9$         \\
$T_{\rm sw}$                  & $10000$K     & $10000$K     & $8000$K      \\
Inclination               & $80^{\circ}$ & $60^{\circ}$ & $60^{\circ}$ \\
\hline
\end{tabular}
\label{tab:examples}
\end{table}

\subsection{Reipurth classification}
\label{sec:reipurth}

Following the framework given by \citet{1996A&AS..120..229R} and outlined in \S~\ref{sec:observations} the synthetic line profiles were classified using a \textsc{Python} routine. The code finds peaks and troughs using a gradient search and uses them to classify the lines with the following order of precedence: \textit{IV}, \textit{I}, \textit{II}, and \textit{III}. The program defines type \textit{IV} as being a sub-continuum absorption with no further emission greater than $1$ per cent of the peak. If the line contains both red and blue-shifted absorption, the code uses the strongest feature to classify the line. Many profiles have type \textit{II}, and \textit{III} features along with type \textit{IV}. To avoid any classification bias, the spectra from X-shooter are classified using the same order of precedence.

The mean signal to noise ratio (SNR) of the X-Shooter spectra are $\approx 70$ for the VIS arm and $\approx 108$ for the NIR. The mean ratio of peak emission to sub-continuum absorption depth for type \textit{IV} profiles is $\approx 3$ and $\approx 4$ for the optical and infrared lines, respectively. Hence, the sub-continuum absorption features in the synthetic grid are $18$ -- $25$ times greater than the observed noise level. Consequently, the introduction of noise would not be expected to significantly influence the Reipurth classifications of the synthetic profiles.

Fig.~\ref{fig:reipurth} displays the results as hashed bars overlaying the distribution of Reipurth types for the X-Shooter spectra. The \ha{} classifications for the models have the most similar distribution to the observations. IPC profiles are seen in the spectra of one of the 29 T Tauri stars (\citet{1996A&AS..120..229R} observed two out of a sample of 43), approximately 3.4 per cent. Whereas, $35$ per cent of the synthetic \ha{} lines exhibit this morphology. The \ha{} IPC profiles are only present in models that would be classified as Weak Line T Tauri stars, based on the \ha{} 10 per cent width criterion of \citet{White:2003gy}. Blue-shifted sub-continuum absorption is relatively rare for \ha{} and only seen in seven per cent of the synthetic profiles.

In contrast to the relative agreement seen for \ha{}, the synthetic infrared lines predict a significantly higher frequency of P-Cygni profiles; $\approx98$ per cent are type \textit{IVR} or \textit{IVB}. However, across all three infrared lines, the mean frequency of IPC profiles in the X-Shooter sample is $\approx23$ per cent and none of the infrared observations exhibit blue-shifted sub-continuum absorption. Comparatively, \citet{2001AA...365...90F} saw a frequency of type \textit{IVR} profiles of 34 and 20 per cent for \pb{} and \pg{}, respectively. Furthermore, they observed no type \textit{IVB} profiles in their sample of 50 T Tauri stars.
 
 For the chosen parameters of our model grid, the synthetic line profiles show a much higher occurrence of deep absorption features than is present in our observed sample. We explore possible modifications to the model in \S~\ref{sec:linebroadening} and discuss them further in \S~\ref{sec:discusion}.  

\subsection{Comparison of widths}
 \label{sec:widths}
 
 \begin{figure}
    \centering
    \includegraphics[width=\linewidth]{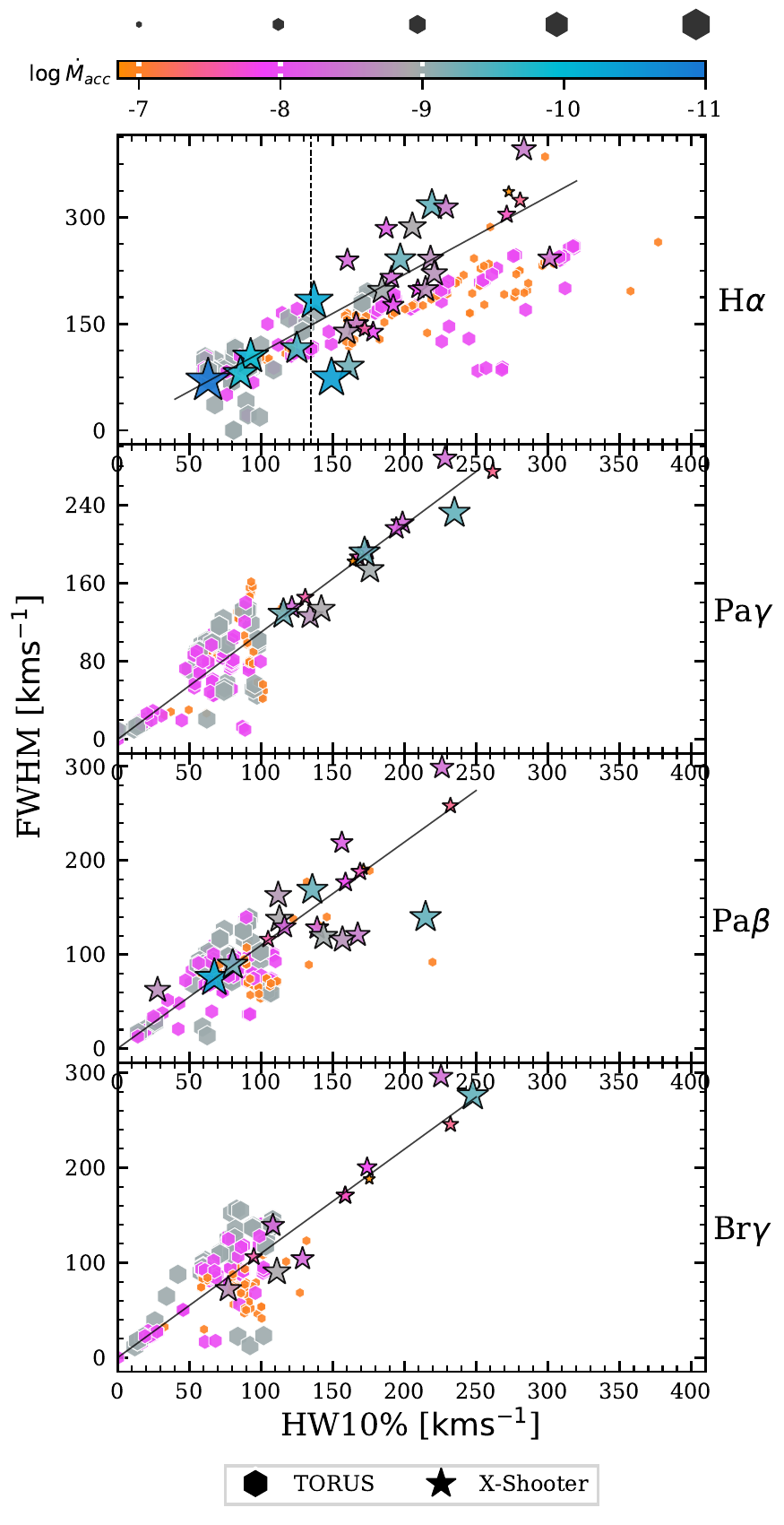}
    \caption{The FWHM vs. HW10\% for observed (black outlined stars) and \textsc{TORUS} data (coloured hexagons). The colour and size denote the accretion rate. The vertical dashed line shows the criterion for the \ha{} split between WTTS and CTTS \citep{White:2003gy}. The diagonal black line is the analytical Gaussian relationship between the FWHM and the HW10\%.}
    \label{fig:widths}
\end{figure}

To analyse the synthetic and observed emission as an ensemble, we calculated the FWHM and the HW10\%. A linear interpolation was fitted to the synthetic line profiles so that the FWHM and the HW10\% could be measured. We determined the FWHM and the HW10\% of the X-Shooter spectra by fitting the line profile with Gaussians rather than a spline to avoid problems with noise and to conform to standard data reduction methods. Two Gaussians were fitted to the emission, and another two Gaussians were used solely to fit the sub-continuum absorption, fitted using a \textsc{Python} routine that prioritised reproducing the observed widths.

Fig.~\ref{fig:widths} shows a plot of the FWHM against the HW10\% for the synthetic and observed spectra. Each point corresponds to a different model in our grid, and the colour and size indicate the accretion rate. The cutoff between Weak Line and Classical T Tauri stars is shown by the dotted vertical line. For \ha{}, FWHM and the HW10\% are strongly correlated for both the observations and models. Both the models and the observed data show that the widths are correlated with the accretion rate. However, the accretion funnel temperature and viewing inclination have significant influence on the width of the model line profiles by scattering in the synthetic line profiles at a given accretion rate. The relationship between the FWHM and HW10\% for a Gaussian is $\approx 1.10$, displayed in Fig.~\ref{fig:widths} by the inclined solid black lines. The observed sample is scattered around this relationship. The synthetic \ha{} spectra have been selected to have a similar range of HW10\% as the observations. However, they tend to have a smaller FWHM for a given HW10\%. The deviation from the Gaussian FWHM to HW10\% relationship is due to Stark broadening. The \ha{} lines have a Voigt profile when Stark broadening is dominant (see \S~\ref{sec:profiles}). The Voigt form has wider wings and a narrower peak than a Gaussian curve. This effect is not observed in the higher lines where Stark broadening is negligible. 

For \pg{} and \bg{}, the observations exhibit a tight correlation between the FWHM and HW10\%. The \pb{} observations exhibit a larger scatter, suggesting a divergence away from the Gaussian form and more structure in the line profiles. This is corroborated by the fact that \pb{} has the highest frequency ($63$ per cent) of non type \textit{I} profiles (see Fig.~\ref{fig:reipurth}). In contrast to \ha{}, the FWHM and the HW10\% of the higher lines in the observations and models exhibit little correlation with the accretion rate. The models shown in Fig.~\ref{fig:widths} have been selected so that the range of HW10\% widths for \ha{} roughly correspond to the range of the observational data set.  However, these models predict widths for the three higher lines that are, on average, much narrower than the observed widths. In general for an accretion rate of $\log{\dot{M}_{\rm acc}}=-7$ the synthetic lines have a lower ratio of FWHM to HW10\% than other accretion rates. 

\begin{figure}
    \centering
    \includegraphics[width=\linewidth]{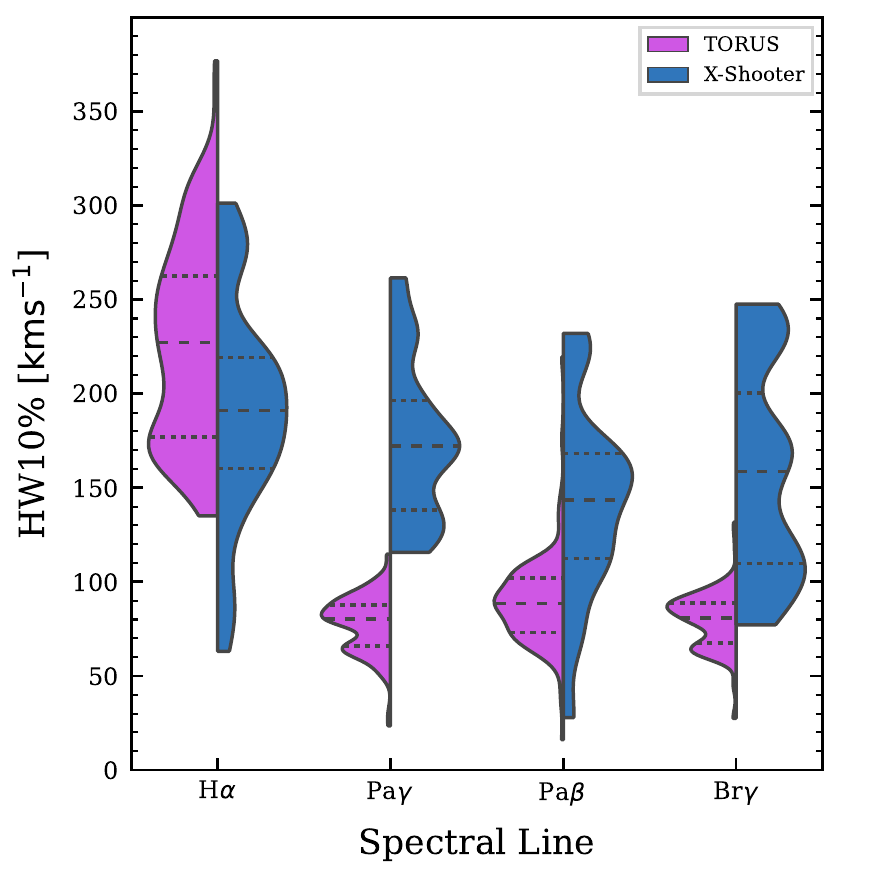}
    \caption{The HW10\% distribution of the grid and observations shown for each spectral line. The model data is filtered so that only models with widths such that $60\leq{\rm~ HW10\%}_{{\rm~ H}\alpha}<400$ are included. The horizontal dashed lines indicate the 25\textsuperscript{th} percentile, the median, and the 75\textsuperscript{th} percentile of the distributions.}
    \label{fig:wisker}
\end{figure}

Fig.~\ref{fig:wisker} compares the distributions of synthetic and observed HW10\% for all four lines. The distribution of synthetic \ha{} widths is much broader than that observed, simply a result of our parameter grid sampling a relatively uniform coverage of line profiles in this range. However, by constraining our models to a subset of the full grid (see \S~\ref{sec:results}) the observed and synthetic \ha{} HW10\%'s lie in a similar range, with a difference in the respective medians of $25{~\rm kms}^{-1}$. Conversely for all three of the other spectral lines, the median width of the synthetic line distribution is much narrower than that in the observed sample. The differences in the median value of HW10\% for the synthetic and observed infrared line profiles are, Pa$\beta = 66{~\rm kms}^{-1}$, Pa$\gamma = 105 {~\rm kms}^{-1}$, and Br$\gamma = 82 {~\rm kms}^{-1}$. In other words, for a range of models that predict \ha{} line widths in the same range as the observations, the predicted widths in the other lines are systematically too narrow by $\approx 84{~\rm kms}^{-1}$. In \S~\ref{sec:linebroadening} we explore what modifications can be made to the models to bring these synthetic lines into better agreement with the observed sample.

\subsection{Modified models}
\label{sec:linebroadening}

The grid of models does not adequately reproduce the observed frequency of line profile structures (\S~\ref{sec:reipurth}) or the observed range of profile widths (\S~\ref{sec:widths}) for the infrared lines. In the following section, we examine the effects of three different model modifications in an attempt to explore what might bring the synthetic and observed line profiles into better agreement. In \S~\ref{sec:turbulent} we examine the effect of introducing turbulent broadening. The influence of rotation on the synthetic lines is introduced in \S~\ref{sec:rotation}. In \S~\ref{sec:blueside} the widths are reevaluated, considering only the blue-shifted side, to determine the consequence of the high proportion of sub-continuum absorption on the widths of the line profiles.
\subsubsection{Turbulent line broadening}
\label{sec:turbulent}

\begin{figure}
    \centering
    \includegraphics[width=\linewidth]{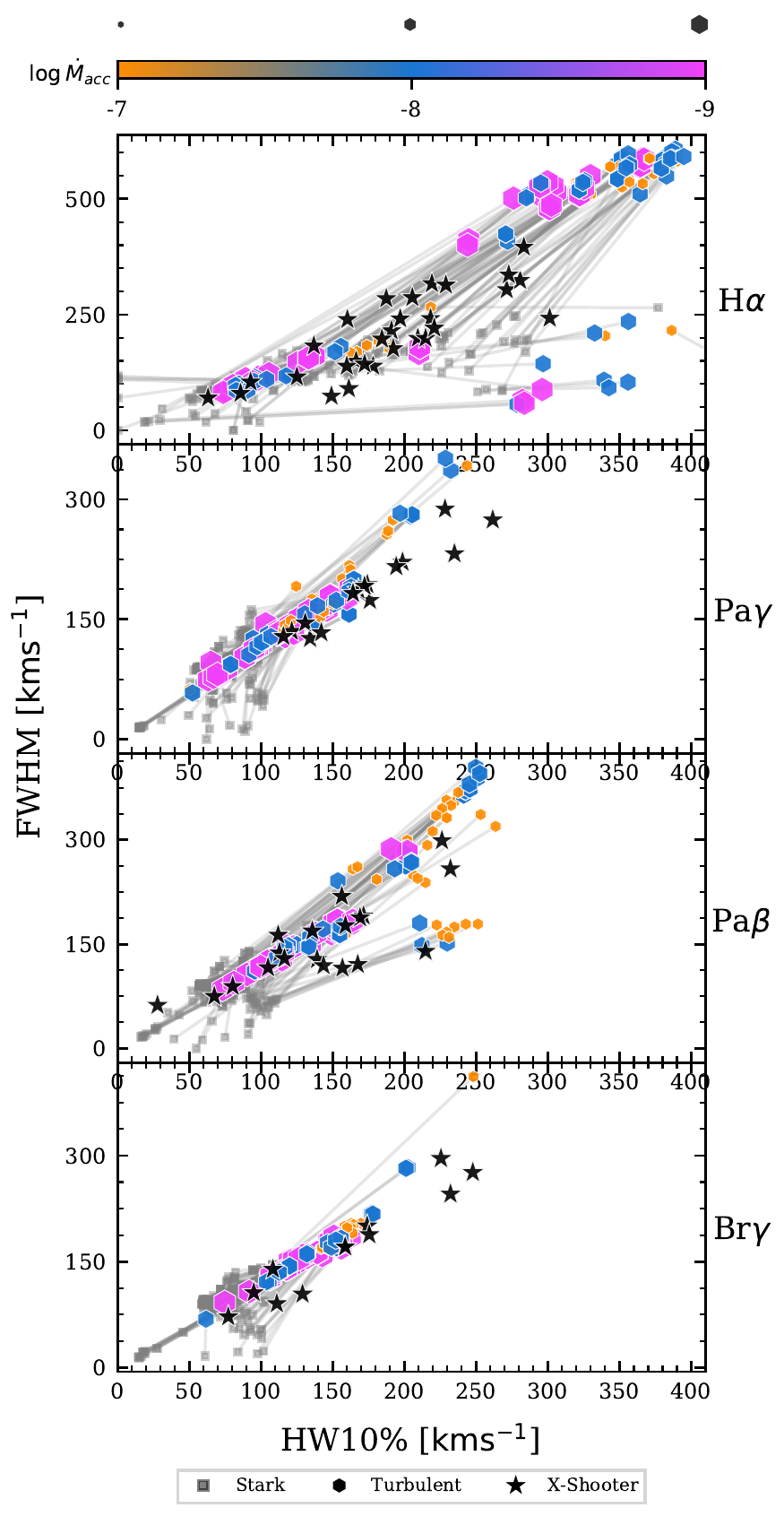}
    \caption{The effect of combining $100{\rm~ kms}^{-1}$ of turbulent velocity with the Stark broadening. Grey squares are the original Stark only broadened widths and the coloured hexagons are the Stark and turbulently broadened widths. The grey lines connect the two different models. The accretion rate of the models is indicated by the colour and size of the hexagons. The black stars indicate the widths of the observed T Tauri stars. The added turbulent velocity increases the width of the synthetic profiles bringing the ensemble into better agreement with the observations.}
    \label{fig:turbwidths}
\end{figure}

We expect there to be unresolved small-scale motion in the T Tauri system on top of the bulk gas flow, for example, from unsteady accretion and wind flows or from instabilities in the accretion and post-accretion shock gas. However, rather than model the physical processes involved, we assume that the unresolved motions are everywhere in the grid and act like a turbulence, which has a Maxwell–Boltzmann distribution. The effect of Stark broadening on the higher lines is negligible, so an additional broadening from turbulence could increase the synthetic line widths to be more in line with the observations and reduce the frequency of sub-continuum absorption features. To explore the effect of turbulent broadening on the infrared line widths, we introduced a large turbulent velocity component to equation~(\ref{eq:doppler}). We used a value of $v_{\rm turb}=100{\rm~ kms}^{-1}$, which is approximately the largest difference of the median HW10\% between the observed and synthetic infrared lines, see \S~\ref{sec:widths}. There is little physical justification for a turbulent motion of $100{\rm~ kms}^{-1}$, other than it is the required velocity to broaden the ensemble of line profiles sufficiently.

Fig.~\ref{fig:turbwidths} shows the FWHM versus the HW10\% for the Stark and turbulently broadened lines and the Stark-only broadened models. Similar to Fig.~\ref{fig:widths}, the profiles are filtered such that only models with an \ha{} HW10\% between $60-400 {\rm~ kms}^{-1}$ are included. The turbulent broadening increases the \ha{} line too much, pushing the median value of the HW10\% to $\approx 320{\rm ~kms}^{-1}$. The effect on the higher lines is to bring the range of predicted values to overlap with the observed spectra. The turbulent broadening increases the number of synthetic lines that have HW10\% greater than $400{\rm~ kms}^{-1}$ by $4$ per cent.

\begin{landscape}
\begin{figure}
    \centering
    \includegraphics[width=0.9\linewidth]{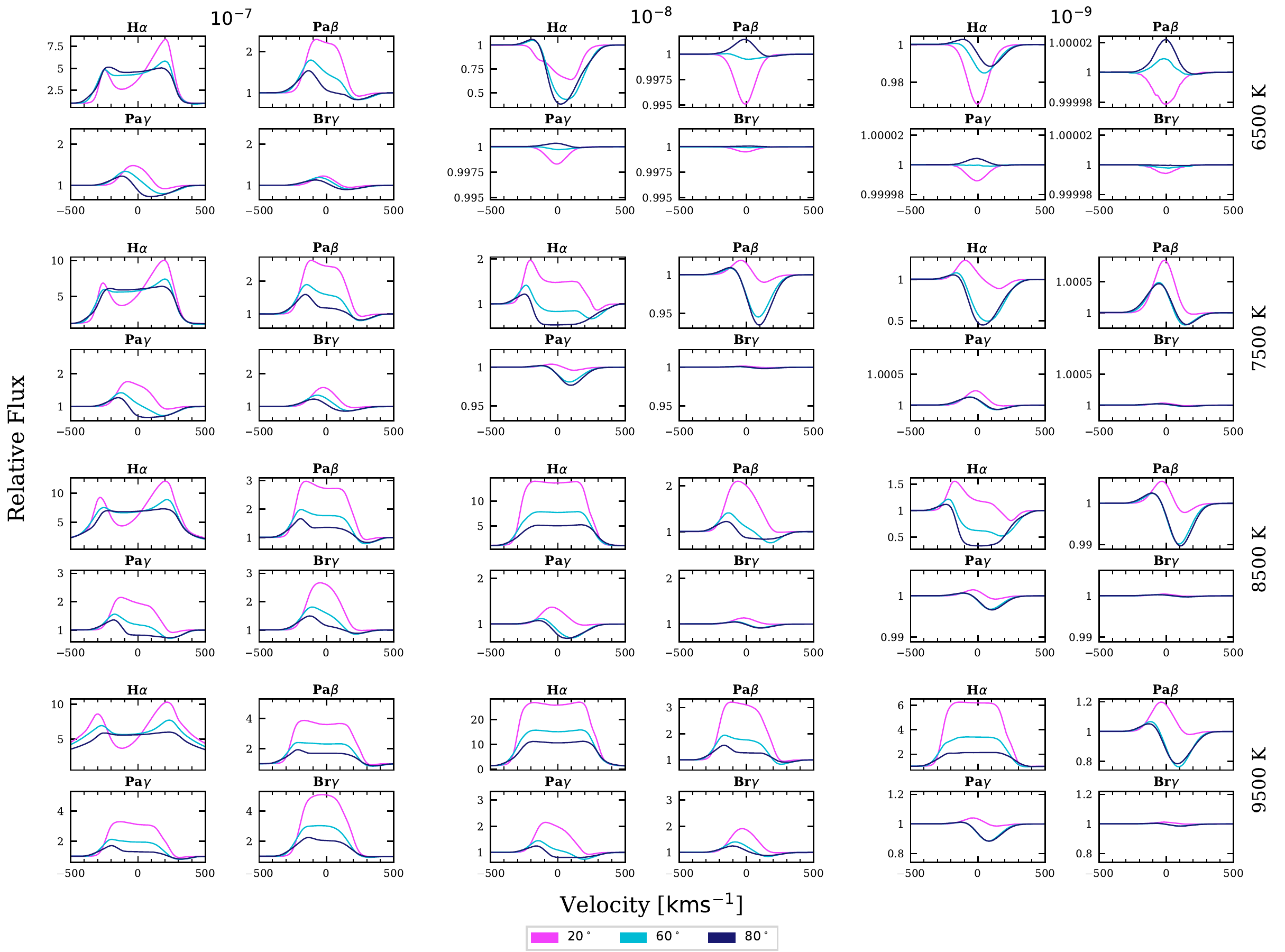}
    \caption{Turbulently broadened \textsc{TORUS} line profiles arranged by $\dot{M}_{\rm acc}$ (columns) and $T_{\rm acc}$ (rows). The y-axis range of \pg{}, \pb{}, and \bg{} for each subset are shared. The profiles are boxy with a ''top-hat'' like appearance and the sub-continuum absorption features are broadened.}
    \label{fig:allModelsTurb}
\end{figure}
\end{landscape}

A subset of the turbulently broadened line profiles can be seen in Fig.~\ref{fig:allModelsTurb}. Similarly to Fig.~\ref{fig:allModels}, the spectra are shown for $\dot{M}_{\rm sw}=0.01~\dot{M}_{\rm acc}$ and $T_{\rm sw}=8000{\rm~K}$. The figure shows the line profiles organized by accretion rate (columns) and magnetospheric temperature (rows). The effect of the turbulence is to create a boxier profile; steep sides with flat tops. The added turbulent velocity increases the resonant range of the transition wavelength, broadening both the emission and the absorption features. This ``top-hat'' morphology is not seen in the observed data set. The distribution of the Reipurth classifications is not significantly changed by the turbulence and a high proportion of IPC profiles are still predicted. The boxier profiles increase the ratio between the FWHM and the HW10\%, creating a steeper relationship than seen for a Gaussian or a Voigt profile.  

In conclusion, despite the fact that the turbulent broadening produces an infrared range of line widths in better agreement with the observations, this modification causes the width of the \ha{} lines to be increased too much and the line profiles to become boxy, a morphology not reflected in the observations. There is also little motivation for the turbulent motion to be as high as $100~{\rm kms}^{-1}$. Furthermore, the frequency of sub-continuum absorption is not significantly changed, suggesting that incorporating unresolved small-scale motion into the simulations cannot solely account for the lack of width in the infrared lines.

\subsubsection{Rotational line broadening}
\label{sec:rotation}

\begin{figure}
    \centering
    \includegraphics[width=0.95\linewidth]{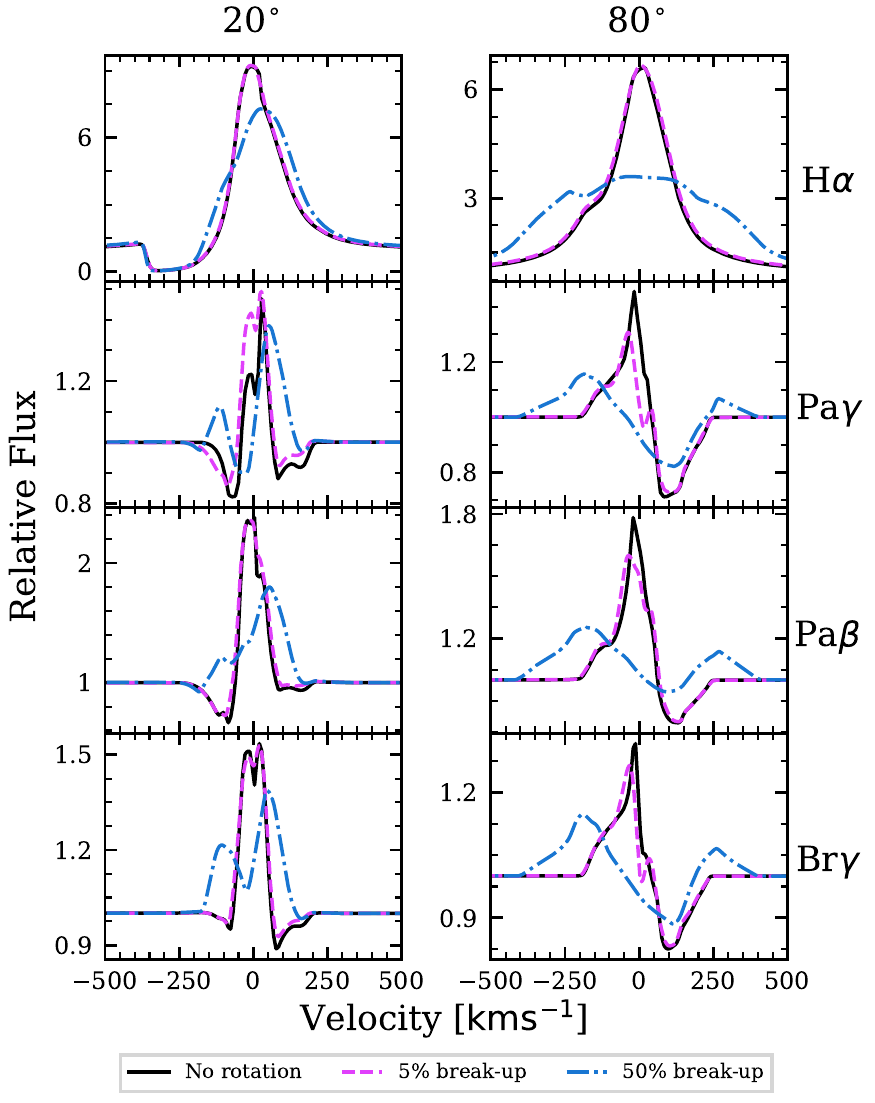}
    \caption{The effect of rotation on line profiles. The profiles are shown for three different rotation rates: no rotation (solid line), $0.05~v_{\rm br}$ (pink dashed line), and $0.5~v_{\rm br}$ (blue dash-dotted line). The profiles are arranged by viewing inclination (columns) and wavelength (rows). At high rotation rates, the profiles are flattened and broadened. The effect is amplified when viewed from a greater inclination.}
    \label{fig:rotation}
\end{figure}

The line profiles presented in \S~\ref{sec:parameterStudy} are computed for a non-rotating T Tauri star model. In principle, the Doppler shift from rotational motion should broaden the line profiles. Thus, to explore whether this improved the model's fit to the data, we ran a small grid of rotating models with a wind velocity as prescribed in \S~\ref{sec:wind}. The magnetosphere is treated as a solid body that rotates with the star. The models were computed at two rotation velocities, $0.05~v_{\rm br}$ ($13.4~{\rm kms}^{-1}$) and $0.5~v_{\rm br}$ ($134~{\rm kms}^{-1}$), where $v_{\rm br}$ is the stellar break-up velocity, taken to be the balance of gravitational and centrifugal forces without considering the effects of an equatorial bulge such that
\begin{equation}
    v_{\rm br}=\sqrt{GMR_{*}^{-1}}{\rm .}
\end{equation}
\noindent
T Tauri stars are typically slow rotators \citep[e.g.][]{2014prpl.conf..433B}. However, we include a model with a rotation rate of $0.5~v_{\rm br}$ so as to be able to compare the effect of an extreme case relative to a more realistic rotation rate.
The models have an accretion rate of $\log\left(\dot{M}_{\rm acc}\right)=-7$, a mass loss rate of $\dot{M}_{\rm sw} = 0.1\dot{M}_{\rm acc}$, a maximum accretion funnel temperature of $T_{\rm acc}=7500{\rm~ K}$, and a stellar wind temperature of $T_{\rm sw}=8000{\rm~ K}$. Fig.~\ref{fig:rotation} shows the line profiles for the rotating and the non-rotating case.

The profiles are smoothed and broadened by rotation. However, this effect is negligible unless the rotation is at a significant proportion of the stellar break-up velocity. For example, a rotation velocity of $0.05~v_{\rm br}$ had a no effect on \ha{}, whereas for \pg{} and \pb{}, the peak intensity was reduced by $\approx10$ per cent for an inclination of $80^{\circ}$. On the other hand, a rotation rate of $0.5~v_{\rm br}$ significantly broadens the synthetic profiles when viewed from a high colatitude. The peaks are shifted away from the rest velocity as the hot, high-density gas in the magnetosphere dominates the emission. Estimates for the observed X-Shooter target's $v\sin i$ and their rotation as a fraction of their break-up velocity are shown in Table \ref{tab:rotation}. V354-Mon has the highest rotation rate in the data set with a velocity of $v\sin{i}=40.9~{\rm kms}^{-1}$, only $11$ per cent of the star's break-up velocity. The rotation rates in our observed data set are not large enough to cause significant line broadening. Therefore, rotational broadening cannot explain the mismatch between the observed and synthetic line widths. 

\begin{table}
\centering
\caption{Table showing the rotation rates for the observed X-Shooter stars. References are: 1 \citet{2012ApJ...745..119N}, 2 \citet{2009AJ....138..963B}, 3 \citet{2015A&A...575A...4F}, 4 \citet{2019AJ....157..196K}, 5 \citet{2014ApJ...788...81M}, 6 \citet{2016A&A...585A.136M}, 7 \citet{2018A&A...614A.108S}, 8 \citet{2014A&A...568A..18M}, 9 \citet{2015A&A...577A..11M}, and 10 \citet{2018A&A...609A..70R}}
\begin{tabular}{lccccc}
\hline
\hline
\multirow{2}{*}{Name} & $L_{*}$   & $R_{*}$   & $V\sin{i}$  &  \multirow{2}{*}{$\frac{V\sin{i}}{V_{br}}$} & \multirow{2}{*}{Reference} \\
     & [${\rm~ L}_{\odot}$] & [${\rm~ R}_{\odot}$]     & [${\rm~ kms}^{-1}$] & & \\
\hline
ESO-Ha-562                   & 0.12       & 0.8            & 40     & 0.1                           & 1,6                     \\
IQ-Tau                       & 0.75       & 1.9            & 14.4   & 0.05                            & 1,6                     \\
V354-Mon                     & 1.67       & 1.8            & 40.9   & 0.11                           & 2,7                     \\
T-33                         & 0.69       & 1.1            & 16.5   & 0.04                          & 3,6                     \\
T-11                         & 1.45       & 1.7            & 14.1   & 0.04                            & 3,8                     \\
VW-Cha                       & 1.64       & 2.6            & 13     & 0.04                            & 1,6                     \\
T-6                          & 1.17       & 1.5            & 34     & 0.09                            & 1,8                     \\
CR-Cha                       & 3.26       & 2.3            & 35     & 0.09                            & 1,6                     \\
T-52                         & 2.55       & 2.0            & 28     & 0.08                            & 1,6                     \\
T-38                         & 0.13       & 0.8            & 18.7   & 0.05                            & 1,6                     \\
KV-Mon                       & $\sim$     & $\sim$         & 24.25  & $\sim$                         & 4,9                     \\
Sz-22                        & 0.51       & 1.3            & $\sim$ & $\sim$                         & 6                       \\
T-4                          & 0.43       & 1.3            & 12.4   & 0.03                            & 1,6                     \\
DG-Tau                       & 1.55       & 2.2            & 24.7   & 0.07                            & 1,6                     \\
T-23                         & 0.32       & 1.8            & 9.6    & 0.05                            & 3,6                     \\
RECX-12                      & 0.23       & 1.4            & 6.4    & 0.03                            & 5,10                    \\
Cha-Ha-6                     & 0.07       & 1.0            & $\sim$ & $\sim$                         & 6                       \\
T-12                         & 0.15       & 1.3            & 10.7   & 0.06                            & 1,6                     \\
CT-Cha-A                     & 1.5        & 2.2            & 7.8    & 0.02                            & 1,6                     \\
ESO-HA-442                   & $\sim$     & $\sim$         & $\sim$ & $\sim$                         &                       \\
CHX18N                       & 1.03       & 1.4            & 26.5   & 0.07                            & 1,6                     \\
TW-Cha                       & 0.38       & 1.2            & 11.3   & 0.03                            & 1,6                     \\
RECX-6                       & 0.1        & 0.9            & $\sim$ & $\sim$                         & 10                      \\
T-49                         & 0.29       & 1.6            & 8.2    & 0.04                            & 1,6                     \\
T-35                         & 0.33       & 1.2            & 21     & 0.05                            & 1,8                     \\
T-45a                        & 0.34       & 1.2            & 12.4   & 0.03                            & 1,6                     \\
RECX-9                       & 0.095      & 1.1            & $\sim$ & $\sim$                         & 10                      \\
T-24                         & 0.4        & 1.4            & 10.5   & 0.03                            & 1,6                     \\
Hn-5                         & 0.05       & 0.8            & 7.8    & 0.04                            & 1,6                    \\
\hline
\end{tabular}
\label{tab:rotation}
\end{table}
\subsubsection{Blue-shifted emission}
\label{sec:blueside}
Over $80$ per cent of the infrared emission lines from the models have IPC features, much more frequent than the $\sim 23$ per cent of observed line profiles with IPC profiles (\S~\ref{sec:observations}). The red-shifted sub-continuum absorption makes the line profiles narrower. Thus, if our model were modified to reduce the frequency of IPC profiles, we expect that the average line widths will also increase. A possible mechanism for removing the red-shifted absorption is introducing a non-axisymmetric accretion flow where the accreting column only periodically intersects with the line of sight. For example, \citet{Esau:2014is} used a non-axisymmetric accretion flow to reproduce the Balmer variations seen for AA Tau successfully; this effect is discussed further in \S~\ref{sec:discusion}.

\begin{figure*}
    \centering
    \includegraphics[width=\linewidth]{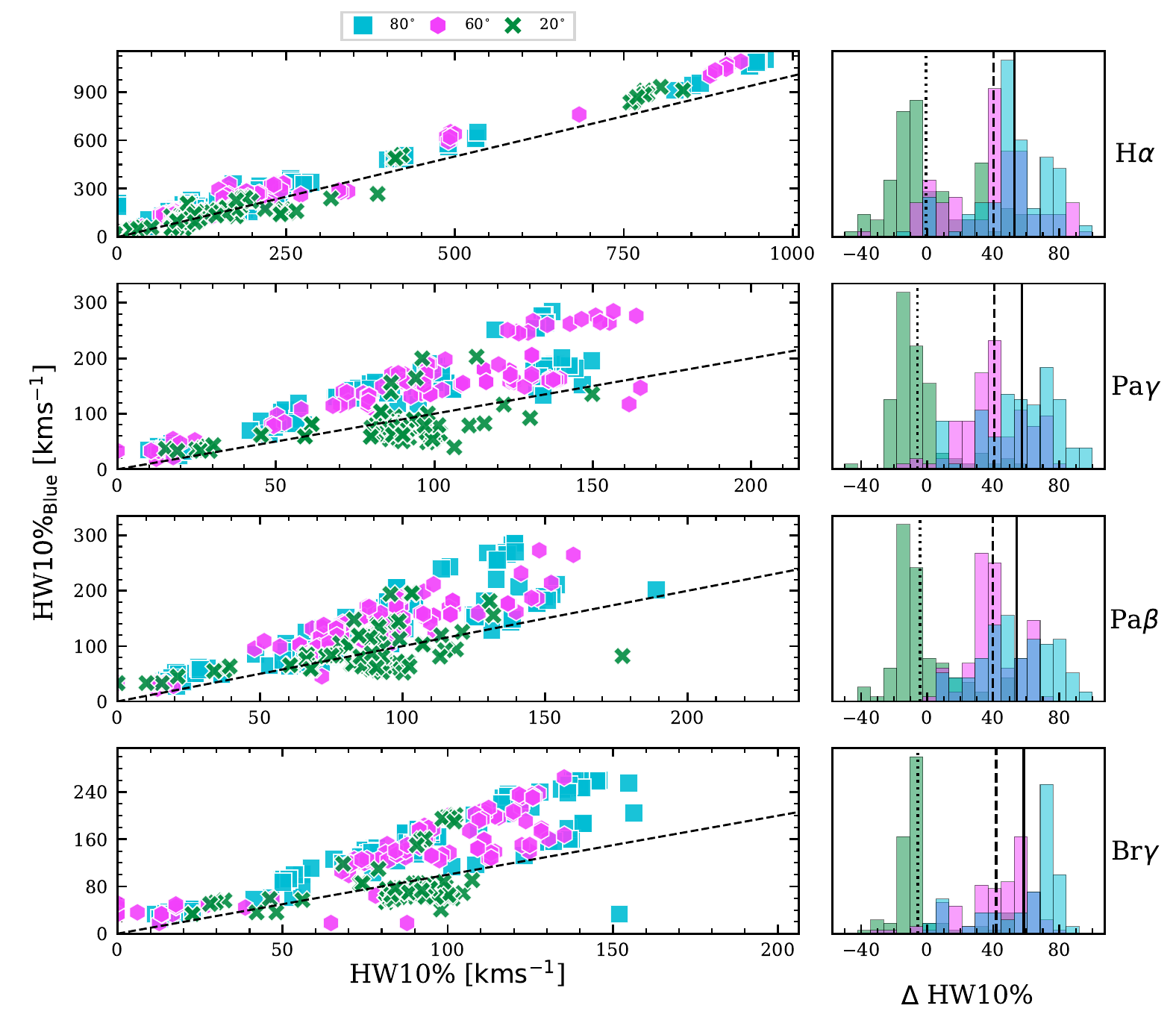}
    \caption{Comparison of the true HW10\% and the half-width measured for only blue-shifted emission (HW10\%$_{\rm Blue}$). The dashed diagonal line has a gradient of unity; points above it are broader when only the blue-shifted emission is considered. The colour and style of the points denote the model inclination. Plotted on the right are histograms showing the difference of the two measured widths. The vertical lines indicate the respective means: $20^\circ$ dotted, $60^\circ$ dashed, and $80^\circ$ solid.}
    \label{fig:blueWidth}
\end{figure*}

To estimate how much the IPC absorption might be affecting the line widths in our models, we calculated the HW10\% values using only the blue side of each synthetic line profile (HW10\%$_{\rm Blue}$). We employed a linear interpolation of the line that was blue shifted from the peak to determine values of the profile width. Fig.~\ref{fig:blueWidth} shows the HW10\%$_{\rm Blue}$ versus the true HW10\%. The diagonal line (black dashed) shows the one to one ratio; points below the diagonal indicate the width decreased when considering only the blue side. The points are coloured, indicating the model's inclination. For a viewing angle of $20^{\circ}$, the majority of line widths decreased when solely considering the blue-shifted emission. The blue sides of these modelled lines are narrower than the red due to the strong blue-shifted absorption from the stellar wind, which lies along the line of sight at this inclination. Type \textit{IV} profiles are rare in our observed data set, and so to focus on the broadening effect they are excluded from the following analysis. Fig.~\ref{fig:blueWidth} shows the distribution of width-change ($\Delta$ HW10\%) as a series of histograms. The mean shift for all the models with an inclination of either $60^{\circ}$ or $80^{\circ}$ is $49{~\rm kms}^{-1}$. When only models with a very low mass-loss rate ($0.001\dot{M}_{\rm acc}$) are considered the average increase in width is $\approx 50{~\rm kms}^{-1}$ for inclinations of $60^{\circ}$ and $80^{\circ}$. 

The above analysis suggests that if the models were modified in some way to reduce the prevalence of IPC absorption, the synthetic line profiles would become significantly wider by an average of $\approx49{~\rm kms}^{-1}$. An average increase of $\approx 84{~\rm kms}^{-1}$ (\S~\ref{sec:widths}) is needed to shift the HW10\% of the synthetic infrared lines into agreement with the observations. Therefore, while this effect cannot account for the whole disparity of widths on its own, when combined with other mechanisms, such as turbulence and rotation, it may be sufficient.

\section{Discussion}
\label{sec:discusion}

A significant caveat to our comparison of the observed and synthetic line profiles is the possibility that our observed data set is incongruous and not representative of the consensus of T Tauri stars. However, our sample of 29 stars appears to be consistent with other observational datasets. For instance, the sample has a similar fraction of Reipurth classes to the distributions reported by \citet{1996A&AS..120..229R} for \ha{} and \citet{2001AA...365...90F} for \pb{} and \bg{}. Additionally, the \pb{} and \bg{} line peaks are slightly blue-shifted in agreement with \citet{2001AA...365...90F}, even though the mean FWHM of our line profiles is smaller than the mean FWHM reported by \citet{2001AA...365...90F}.

A second caveat is that our grid of synthetic line profiles were not adjusted to attain an agreement with any particular observation and a singular stellar mass, radius, and temperature were used. Broadly speaking, the synthetic line profile morphology is determined by the geometry, accretion rate, and temperature of the magnetosphere. Accordingly, our grid was designed to encompass the broad range of parameters predicted for T Tauri accretion and outflow \citep[e.g.][]{Hartmann:1994tl,1998ApJ...492..743M,2001ApJ...550..944M,Kurosawa:2006gd,Lima:2010bo,2012MNRAS.426.2901K}. In turn, our T Tauri sample has an extensive range of accretion rates and stellar parameters, making it an appropriate initial dataset for a comparison with the ensemble of synthetic spectra.

Our models predict a greater number of \ha{} IPC profiles than observed. However, our grid of models can reproduce the observed fraction of \ha{} Reipurth types, after the lines classified as Weak Line T Tauri stars ($\approx 40$ per cent) are removed. By removing the Weak Line T Tauri models, we discard the \ha{} models that contain red-shifted sub-continuum absorption. Hence, for our model grid, IPC profiles are only produced by models that would be classified as Weak Line T Tauri stars. This result is substantiated by \citet{Kurosawa:2006gd} who demonstrated that \ha{} IPC profiles were only produced in $\approx5$ per cent of their sample. Moreover, in these cases, the accretion rate was low enough that the lines would be consistent with a weak line T Tauri classification.

The same models that correctly reproduce the observed \ha{} Reipurth types predict a very high frequency of IPC profiles for \pb{}, \pg{}, and \bg{}. For \ha{}, Stark broadening smooths out the synthetic spectra, masking absorption features \citep{2001ApJ...550..944M}. However, for the higher lines, the effect of Stark broadening is negligible at the temperatures and densities expected in T Tauri accretion. A possible reason for the dearth of observed IPC profiles is non-axisymmetric accretion flow. A magnetic obliquity from the stellar pole could cause the accretion funnel to form ``curtains'' that rotate with the star. IPC profiles would then be seen intermittently when the accretion curtain intersects with the line of sight. Observations support this effect; for example, \citet{2020MNRAS.497.2142M} measured a magnetic obliquity of $18^{\circ}{}^{+8}_{-7}$ for DK Tau and observed an IPC profile for two of the eight epochs. Accretion funnel ``curtains'' have also been explored in radiative transfer studies. For instance, \citet{2005MNRAS.356.1489S}  used a 3D non-axisymmetric model, to reproduce some of the observed variability seen in T Tauri stars, but in general, the models produced results that were too variable. \citet{Esau:2014is} successfully modelled variable Balmer emission from AA Tau with a non-axisymmetric accretion model and \cite{2012A&A...541A.116A} reproduced Balmer lines from V2129 Oph over several stellar rotational cycles. Further analysis of T Tauri time-series spectroscopic surveys would be helpful in better quantifying the variability of IPC profiles across multiple wavelengths.

In addition to the red-shifted features from the magnetosphere, some interesting blue-shifted detail are created by the stellar wind. The stellar wind produces the characteristic broad sub-continuum absorption as noted by \citet{Kurosawa:2011fh}. P-Cygni features occur in $\approx 13$ per cent of the synthetic profiles, and $\approx 10$ per cent of those do not contain red-shifted sub-continuum absorption. On the other hand, P-Cygni profiles are very rare in our observations, with none exhibited for the infrared lines. The infrared lines are optically thinner than \ha{}, and therefore, less sensitive to the lower density wind. P-Cygni profiles were also noted to be rare by \citet{2001AA...365...90F} and \citet{2006ApJ...646..319E}. Why these mass outflow signatures are scarce in observations could be a matter of geometry, temperature, or mass loss rate. Geometry should significantly reduce the frequency of fast wind absorption, since those winds are likely to be collimated \citep{2006ApJ...646..319E}. \citet{Lima:2010bo} showed that for \ha{} a mass-loss rate from the inner disc of at least $10^{-9}~M_{\odot}{\rm~yr}^{-1}$ was needed for blue-shifted absorption features to show. However, for a polar stellar wind, our models produced \ha{} blue-shifted sub-continuum absorption for mass-loss rates as low as $10^{-12}~M_{\odot}{\rm~ yr}^{-1}$. Additionally, our results suggest that the frequency of synthetic P-Cygni profiles depends on the ratio of the mass loss to mass accretion rate, not the mass loss rate alone, because higher accretion rates increase the magnetospheric and hot-spot emission which masks the absorption from the wind. 

Another factor seen in \S~\ref{sec:widths}, is that the infrared line widths (FWHM and HW10\%) of the synthetic and observed data sets do not agree. Models that produce \ha{} profiles that lie within the observed range create Paschen and Brackett spectra that are too narrow by $\approx 84{\rm kms}^{-1}$. \citet{2000AJ....119.1881A} suggested that rotation and turbulence may play an important role in the formation of the infrared line profiles. In agreement with \citet{2001ApJ...550..944M} our results demonstrated that stellar rotation has a negligible effect on the line widths at the rotation rates seen in our observed sample. Furthermore, the level of turbulent motion needed to broaden the synthetic lines sufficiently to account for the difference in width of the synthetic infrared and \ha{} lines is high, $\approx 100{~\rm kms}^{-1}$. Not only is there no obvious physical justification for such a large value of turbulence, but the spectra produced are broad and flat-topped, a morphology not seen in observations.

In addition to turbulence and rotation, we attempted to quantify the reduction in the width of the spectra due to red-shifted absorption, by calculating the width using only the blue-shifted emission; an approach similar to that adopted by \citet{2012A&A...541A.116A} to compensate for poorly fitting red-shifted absorption. As expected \citep{2007prpl.conf..479B,2012A&A...541A.116A}, our analysis suggested that a significant change in the model geometry that removed the IPC absorption, such as non-axisymmetric accretion, could account for a substantial increase in the infrared line widths. Alone, such a method may not be sufficient to account for the disparity between synthetic and observed hydrogen profiles. However, if the models were to include some combination of stellar rotation, turbulence and non-axisymmetric accretion, the synthetic profiles may be shifted into accord with the observations. Conversely, \citet{2001AA...365...90F} showed that, for their sample of \pb{} and \bg{} T Tauri spectra, the FWHM of Reipurth type \textit{I} and type \textit{IVR} profiles were not significantly different. This would indicate that spectra with inverse P-Cyni profiles are generally broader than those without or that the sub-continuum absorption does not significantly narrow the spectra. However, there is not a sufficient population of type \textit{IVR}'s in our sample to confirm this result.

Another factor not taken into account is the line emission from the accretion shock region. This work treats the accretion shock zone as a black body heated by the release of gravitational potential energy that adds to the continuum. However, the accretion regions will also have line emission \citep[e.g.][]{2018MNRAS.475.4367D} that will contribute to the observed spectra. A complete radiative equilibrium treatment of the shock region and magnetosphere is necessary to constrain this effect fully, which requires the development of a self-consistent theory of the heating and cooling mechanisms in the magnetosphere. 

\section{Conclusions}
\label{sec:conclusions}

We presented a grid of synthetic T Tauri line profiles, computed using the radiative transfer code \textsc{TORUS}. Our models used the magnetospheric model originally developed by \citet{Hartmann:1994tl} and further developed by \citet{1998ApJ...492..743M,2001ApJ...550..944M,2005MNRAS.356.1489S,Kurosawa:2006gd,Lima:2010bo,Kurosawa:2011fh}. Our models also included a polar stellar wind launched from the star's surface, a more realistic development to stellar wind used by \citet{Kurosawa:2011fh}. We compared our ensemble of models to the spectra of 29 T Tauri stars. The medium-resolution spectra covered a broad range of wavelengths that allowed us to simultaneously compare \ha{}, \pb{}, \pg{}, and \bg{} lines. We drew the following conclusions:
\begin{itemize}[labelindent=2pt,itemindent=0pt,leftmargin=*]

    \item The observations exhibited similar Reipurth classifications, widths, and mean line centres to those seen in other studies \citep[e.g.][]{2001AA...365...90F,1996A&AS..120..229R,2006ApJ...646..319E}.
    
    \item For \ha{}, our grid of synthetic line profiles were able to reproduce the observed distribution of Reipurth classifications and line widths. However, for the same models the infrared lines could not reproduce the observed emission. The modelled lines were too narrow by $\approx 84{~\rm kms}^{-1}$ and the majority ($\approx 90$ per cent) had IPC profiles. 

    \item We explored the effect of rotation and turbulence and determined that they could not sufficiently broaden the synthetic infrared line profiles or remove the IPC features. The red-shifted sub-continuum absorption was determined to narrow the HW10\% of the lines by $\approx49{\rm kms}^{-1}$.
    
    \item The polar stellar wind produced the characteristic broad P-Cygni profiles as suggested by \citet{Kurosawa:2011fh}. Our models produced blue-shifted sub-continuum absorption with mass loss-rates as low as $10^{-12}~M_{\odot}{\rm~ yr}^{-1}$. Furthermore, our results suggested that the frequency of P-Cygni profiles depends on the ratio of the mass loss to mass accretion rate; the increased emission from higher accretion rates masks the blue-shifted absorption. 
\end{itemize}

Looking forward, future radiative transfer models should include 3D non axi-symmetric models and work to obtain a better match to the observed red-shifted absorption and emission features. The radiative transfer models should be compared to higher resolution observational spectra and include a more complete analysis that takes into account for veiling and extinction effects. Furthermore, the radiative transfer models of T Tauri stars contain a plethora of free parameters. The addition of further constraints on the geometry and temperature of these models is vital if radiative transfer models are to be used as a comprehensive diagnostic tool of T Tauri spectra.

\section*{Acknowledgements}
The Authors acknowledge funding from the European Research Council (ERC) under the European Union’s Horizon 2020 research and innovation program (grant agreement No. 682393; AWESoMeStars: Accretion, Winds, and Evolution of Spins and Magnetism of Stars; \url{http://empslocal.ex.ac.uk/AWESoMeStars}).

This research has made use of the services of the ESO Science Archive Facility. Our research is based on observations collected at the European Southern Observatory under ESO programme 084.C-1095(A).

The authors would like to acknowledge the use of the University of Exeter High-Performance Computing (HPC) facility in carrying out this work.

This research has made use of the SIMBAD database,
operated at CDS, Strasbourg, France \citep{2000A&AS..143....9W}

We thank the developers of the Python packages, Pandas \citep{jeff_reback_2020_3715232} and Matplotlib \citep{Hunter:2007} that were used for data analysis and creating the figures in this work. 

We thank the anonymous referee who provided us with comments and suggestions that improved the clarity of
the manuscript.

\section*{Data Availability}
The VLT/X-Shooter observations are available in the ESO Science Archive Facility at \url{https://archive.eso.org/scienceportal/home} under the program ID 084.C-1095(A). The radiative transfer data underlying this article was generated using the code \textsc{TORUS} which can be found at \url{http://www.astro.ex.ac.uk/people/th2/torus_html/homepage.html}.


\bibliographystyle{mnras}
\bibliography{library.bib}

\bsp	
\label{lastpage}
\end{document}